\documentclass[preprint,sort&compress,a4paper,10pt]{elsarticle}
\usepackage[margin = 2.8cm]{geometry}

\usepackage{graphicx}
\usepackage{subfigure}
\usepackage{amsmath}
\usepackage{amssymb}
\usepackage{bbold}
\usepackage{mathptmx}
\usepackage{slashed}
\usepackage{bm}
\usepackage[utf8]{inputenc}
\usepackage{tikz}
\usepackage{setspace}

\newcommand{\GeV}{\,\mathrm{GeV}}
\newcommand{\MeV}{\,\mathrm{MeV}}

\newcommand{\fm}{\,\mathrm{fm}}

\newcommand{\Lag}{\mathcal{L}}
\newcommand{\LSM}{\text{QMM}}
\newcommand{\eLSM}{{\text{e}\LSM}}

\newcommand{\Lkm}{\Lag_\text{km}}
\newcommand{\Dqqsp}{\mathcal{D}\overline{\psi}\mathcal{D}\psi\mathcal{D}\sigma\mathcal{D}\vec{\pi}}
\newcommand{\Dsp}{\mathcal{D}\sigma\mathcal{D}\vec{\pi}}

\newcommand{\ie}{\textit{i.e.\,}}
\newcommand{\eg}{\textit{e.g.\,}}
\newcommand{\cf}{\textit{cf.\,}}
\newcommand{\Ord}[1]{\mathcal{O}\left(#1\right)}
\newcommand{\eff}{\text{eff}}
\newcommand{\qq}{\psi}
\newcommand{\diff}[2]{\frac{\partial #1}{\partial #2}}
\newcommand{\variation}[2]{\frac{\delta #1}{\delta #2}}
\newcommand{\vpi}{\boldsymbol{\pi}}
\newcommand{\variance}{\mathrm{var}\,}

\renewcommand{\varpi}{\left\langle \vpi^2 \right\rangle}
\renewcommand{\Im}{\mathrm{Im}\,}

\newcommand{\Tr}{\mathrm{Tr}}
\newcommand{\ret}{\text{ret}}
\newcommand{\fix}{\text{fix}}
\newcommand{\vac}{\text{vac}}
\newcommand{\SourcesSpqqA}{\eta_\sigma,\vec\eta_\pi,\eta_q,\bar\eta_q,\eta_\gamma^\mu}
\newcommand{\Gph}[2][{}]{{\left(G^0_\gamma\right)_{#2}#1}} 
\newcommand{\Gqu}[1][{}]{{\left(G^0_\psi\right)}_{\sigma,\pi}#1} 
\newcommand{\Gquvev}[1][{}]{{\left(G^0_\psi\right)}_{\langle\sigma\rangle,0}#1} 
\newcommand{\Gsi}[1][{}]{{G_\sigma}#1} 
\newcommand{\Gpi}[1][{}]{{G_\pi}#1} 
\newcommand{\dGqu}[1][{}]{{\widehat{G^0_\psi}}#1}
\newcommand{\dGph}[2][{}]{{\left(\widehat{G^0_\gamma}\right)_{#2}#1}}
\newcommand{\dJem}[2][{}]{{\widehat{J_\gamma}^{#2}#1}}
\newcommand{\mNuVac}{{m_\text{nuc}^\text{vac}}}
\newcommand{\mSiVac}{{m_\sigma^\text{vac}}}
\newcommand{\mPiVac}{{m_\pi^\text{vac}}}
\newcommand{\vSiVac}{{\langle\sigma\rangle_\text{vac}}}
\newcommand{\CEP}{\text{CEP}}

\newcommand{\dkdx}{{d^3\!kd^4\!x}}
\newcommand{\Lagbar}[1][{-0.9em}]{\overline{\vphantom{\Lag}\hphantom{L}}\hspace{#1} \Lag}
\renewcommand{\bar}{\overline}

\begin{document}
\title{Photon emissivity in the vicinity of a critical point - A case study within the quark meson model}
\author[HZDR,TUD]{F. Wunderlich\corref{cor}}
\ead{f.wunderlich@hzdr.de}
\author[HZDR,TUD]{B. K\"ampfer}
\ead{kaempfer@hzdr.de}

\address[HZDR]{Helmholtz-Zentrum Dresden-Rossendorf, Institute of Radiation Physics,
              \mbox{Bautzner Landstr. 400}, D-01328 Dresden, Germany}
\address[TUD]{Institut f\"ur Theoretische Physik, Technische Universit\"at Dresden,
              D-01062 Dresden, Germany}
\cortext[cor]{Corresponding author}
\date{\today}
\begin{abstract}
          The quark meson (linear sigma) model with linearized fluctuations displays at a critical end point 
          the onset of a curve of first-order phase 
          transitions (FOPTs) located at non-zero
          chemical potentials and temperatures below a certain cross-over temperature. 
          The model qualifies well for an 
          illustrative example to study the impact of the emerging FOPT, \eg on photon emissivities. Such a case
          study unravels the tight interlocking of the phase structure with the emission rates, here calculated according to
          lowest-order tree level processes by kinetic theory expressions.
          It is the strong dependence of the rates on the effective masses of the involved degrees of freedom which 
          distinctively vary over the phase diagram thus shaping the emissivity accordingly.
          At the same time, thermodynamic properties of the medium are linked decisively to these effective masses, \ie a 
          consistent evaluation of thermodynamics, governing for instance adiabatic expansion paths, and emission rates
          is maintained within such an approach. 
\end{abstract}
          
\maketitle

\section{Introduction}
After several decades of dedicated research, the phase diagram of QCD has revealed a number of fairly 
intricate properties. At zero baryo-chemical potential $\mu_B$ the Columbia plot 
(\cf \cite{Ding:2015ona,Philipsen:2015eya} for recent versions) points to a first ($m_s<m_s^\text{tric}$)
or second ($m_s\geq m_s^\text{tric}$) order phase transition in the chiral limit for the light quark flavors, 
with the position of the tricritical point
$m_s^\text{tric}$ not yet settled, and to a crossover for physical quark masses when considering three quark
flavors with the two light flavors being degenerate.
In this way of thinking the case with all quark flavors set to infinity corresponds to pure gauge theory with
a first-order phase transition (FOPT) at $T_c = \Ord{270\MeV}$. 
Leaving the flavor number and quark mass dependence of $T_c$ (either 
the cross over temperature scale or the critical temperature) to future investigations, much progress
has been achieved for the relevant case with physical quark masses: $T_c = 154\pm8\MeV$ is now the settled continuum 
extrapolated cross over temperature \cite{Aoki:2006br,Bazavov:2011nk}, 
where the description in terms of hadronic (quasi-particle) degrees of freedom
has to be changed in favor of quark-gluon type degrees of freedom.
Much less is known when allowing for non-zero baryo- (and maybe other) chemical potentials. Several techniques
have been developed to access the region $\mu_B/T \lesssim 1$ 
\cite{Fodor:2001au,deForcrand:2002ci,Fodor:2007vv,Kaczmarek:2011zz}.
A non-zero baryo-chemical potential $\mu_B$ is particularly intriguing as the cross over is expected to turn into
a FOPT when moving to larger values of $\mu_B$ \cite{Stephanov:2004wx}. The onset
can be related to a critical end point ($\CEP$) with presently rather uncertain coordinates $(T_\CEP, \mu_\CEP)$.
Such an option of a CEP in the QCD phase diagram has triggered a lot of dedicated activities,
both experimentally \cite{Anticic:2009pe,Mohanty:2009vb,Lacey:2014wqa,Czopowicz:2015mfa} 
and theoretically, applying lattice techniques \eg reweighting \cite{Fodor:2001au}, 
Taylor expansion in $\mu_B$ \cite{Kaczmarek:2011zz},
analytic continuation from imaginary $\mu_B$ \cite{deForcrand:2002ci} or density of state methods \cite{Fodor:2007vv}
as well as Dyson-Schwinger \cite{Fischer:2014ata},
chiral model \cite{Scavenius:2000qd,Schaefer:2004en,Schaefer:2007pw,Herbst:2010rf,Fukushima:2013rx,Kovacs:2016juc} 
or quasiparticle aproaches \cite{Bluhm:2006av} giving widespread results \cite{Stephanov:2004wx}. 
One signature that is looked for is an unsteady behavior of event-by-event 
fluctuations of conserved quantities \eg baryon number or strangeness \cite{Stephanov:1999zu,Asakawa:2000wh} and 
deviations from a Gaussian distribution of these parametrized by higher moments 
such as skewness and kurtosis \cite{Vovchenko:2015pya}.
An overview over possible approaches can be found in \cite{Gupta:2009mu}.

From the experimental side, there exist restrictions originating from astrophysical observations \cite{Klahn:2006ir},
nuclear physics and heavy-ion collisions (HICs).
In the latter experiments, nuclei and protons in various combinations are brought to collision at relativistic energies
and create a system of strongly 
interacting particles which expands rapidly and eventually fragments into hadrons. 
At RHIC and LHC energies, the hot medium produced initially is dense enough to be described in terms of 
nearly ideal relativistic hydrodynamics \cite{Teaney:2000cw}
leading to the notion that the quark-gluon medium is strongly coupled. 
By tuning the collision parameters (\eg beam energy, centrality, system size etc.) 
the strongly interacting medium 
evolves through different parts of the phase diagram and thus peculiarities, 
such as phase boundaries and critical points, may leave imprints in the data.
Transport \cite{Bratkovskaya:2004kv} as well as hydrodynamical \cite{Ivanov:2005yw} calculations show indeed 
that the medium evolves through the region
where chiral and confinement transitions presumably take place.

One tool for investigating the transiently hot and dense medium is provided by hadronic probes. Due to their strong interaction with the
ambient medium they quickly loose the information of the conditions under which they where produced, but they can 
be used on the other hand to probe collective phenomena (\eg elliptic and higher order flow components, jet quenching) 
and transport coefficients \cite{Romatschke:2007mq}
as well as issues concerning strangeness production and quarkonia spectra. 
To obtain information from the hot and dense interior of the ``fireball'' created in HICs and proton-nucleus as well as 
high multiplicity proton-proton collisions, 
weakly interacting probes like photons or dileptons 
provide interesting tools
(see \cite{Gale:2009gc,Rapp:2009yu,Vujanovic:2013jpa,Shen:2013vja,Lee:2014pwa} for recent reviews and further references).
Due to their penetrating nature they monitor all stages in the course of such collisions rendering them very promising but rather 
difficult to analyze. While modern transport codes, \eg UrQMD \cite{Bass:1998ca} or PHSD \cite{Cassing:2009vt}, 
try to calculate the photon (real and
virtual) yields from several or even all states (pre-equilibrium photons, thermal photons from the hydrodynamical stage
and photons from final state decays) \cite{Bauchle:2010ym,Linnyk:2013wma}, other approaches focus on the thermalized stage
applying kinetic theory \cite{Liu:2007zzw,vanHees:2011vb} or relate photon emission to the vector meson current
via the assumption of vector meson dominance \cite{Turbide:2003si}.
A detailed survey on the experimental status can be found in \cite{Tserruya:2009zt}.

Recent analyses \cite{Rapp:2014qha,Gale:2014dfa} reveal a tension between the photon-$v_2$ measurements (pointing to late emission, when the medium
anisotropy has built up) and the $p_T$ systematics (pointing to earlier emission, when the medium is hotter).
As a solution to this puzzle the authors in \cite{Rapp:2014qha,vanHees:2014ida} suggest a ``critical enhancement'' and the 
``semi-quark-gluon plasma''; another option is explored in \cite{Vovchenko:2016ijt}.

For macroscopic systems an otherwise transparent medium becomes opaque near or at a critical point. This critical 
opalescence is a consequence of fluctuations on all length scales which is typically phrased as the correlation length
becoming divergent at criticality \cite{Lesne:1998,Antoniou:2006zb}. Guided by such a phenomenon, one can ask whether 
an equivalent effect may emerge also in a strongly interacting medium. In fact, in \cite{Delorme:1980dle} such a 
possibility has been discussed in the context of pion condensation in compressed nuclear matter.
Having in mind, however, HICs with rapid expansion dynamics, specific features of the photon emissivity, when passing 
nearby or through a critical point, are expected to be masked by pre- and post-critical contributions to the 
time integrated emission rate. We, therefore, focus here on the question whether the CEP-related FOPT
causes peculiarities of the photon emission; in the best case such peculiarities could be strong enough 
that they show up even in integrated rates.
Since our understanding of QCD in the region of interest is in its infancies, as mentioned above, we have to resort
to a model which mimics at least a few of the desired effects, especially the onset of a FOPT.
In the present study we select a special quark-meson model and account for linearized meson fluctuations. That is
quarks and mesons contribute both to the pressure (contrary to the mean field approximation which, in the context of
this model, discards the mesonic fluctuation contributions) thus allowing for a treatment within effective kinetic theory to 
calculate photon production by the respective quasi-particle modes. 

Our paper is organized as follows. In Section~\ref{sec:model} we define the employed quark-meson model 
including the  electromagnetic
sector. In Section~\ref{sec:GenFunc} we derive an approximation of the generating functional for correlation functions,
which is used in Section~\ref{sec:PhotonProp} to derive an expression for the leading order contributions of the 
photon production rate.
The photon spectra are evaluated and discussed in Section~\ref{sec:Rates_eval}. In the spirit of a crosscheck we give in 
\ref{apdx_ThDyn} expressions for the
grand canonical potential, which is tightly related to the generating functional and compare the thermodynamic potential 
with literature. Finally, we summarize in 
Section~\ref{sec:Summary} after a brief discussion Section~\ref{sec:Discussion}. 
Formal manipulations
needed for the propagators as well as the derivative expansion of the fermion determinant are explained in 
\ref{apdx_deriv_exp} and \ref{apdx_invert_mtrx}. 
Some further details of the thermodynamics of the employed model are relegated to \ref{apdx_ThDyn}. 
The squared matrix elements employed in the calculation
of the photon rates are listed in \ref{apdx_mtrx_elem}.

\section{Model definition}
\label{sec:model}
To answer the question to which extent the photon signal can reflect the peculiarities of a phase diagram with a FOPT
we resort to a specific model containing 
fermionic (``quark'') and bosonic (``meson'') degrees of freedom. Its Lagrangian reads ($\tau_i$ are the Pauli matrices)
\begin{align}
   \Lag     &= \Lag_\qq + \Lag_\text{km} - U(\sigma,\vec \pi), \label{Lagrangian_01}\\
   \Lag_\qq &= N_c\bar \psi (i\slashed \partial - g(\sigma + i\gamma^5\vec \tau \vec \pi)) \psi 
            \ \equiv\  N_c\bar\psi \Gqu^{-1}\psi,\label{Lagrangian_02}\\
   \Lag_\text{km}   &= \frac12 \partial_\mu \sigma \partial^\mu \sigma + \frac12 \partial_\mu \vec \pi \partial^\mu \vec \pi,\label{Lagrangian_03}\\
   U        &= \frac\lambda4 (\sigma^2+\vec\pi^2-\zeta)^2 - H\sigma\label{Lagrangian_04},
\end{align}
where we define the operator $\Gqu^{-1} = (i\slashed \partial - g(\sigma + i\gamma^5\vec \tau \vec \pi))$
at given meson fields. The field $\psi$ is a doublet of fermion fields with degeneracy $N_c$
(which is interpreted in the context of strong interactions as the number of colors), 
while $\sigma$ and $\vec\pi$ are iso-scalar and iso-vector spin-0 fields. 
The parameters $g$ and $\lambda$ characterize the strength of the 
fermion-meson coupling and the sigma-pion coupling, respectively, while the two other parameters are measures for the 
explicit ($H$) and spontaneous ($\zeta$) breaking of chiral symmetry.
In the literature, the model is often called the quark meson model ($\LSM$) or the linear sigma model, although
the latter notion sometimes only refers to the purely mesonic model with the Lagrangian $\Lkm-U$, 
which we call the $O(4)$ model.

With the field content just described, the low temperature properties do not agree well with observations at nuclear
density or from neutron stars \eg the pressure at low temperature is too small \cite{Steinheimer:2013xxa}. 
A systematic improvement can be achieved by including axial and vector mesons \cite{Parganlija:2010fz},
enlarging the flavor space to three \cite{Parganlija:2012fy} or even four \cite{Eshraim:2014eka,Sasaki:2014asa}
dimensions and including various symmetry breaking terms for fitting 
vacuum properties of the model to QCD results or meson properties \cite{Parganlija:2010fz}.
The interaction with the gluon field can be partially included by coupling the $\LSM$ to a Polyakov loop for which
several choices for the potential are possible \cite{Schaefer:2007pw,Herbst:2010rf}, \cf also \cite{Stiele:2016cfs}. 
Sometimes even glueballs are introduced \cite{Sasaki:2011sd}.
The $\LSM$ as well as its various extensions are used as tools for mimicking (de-)confinement, chiral symmetry breaking
and restoration \cite{Herbst:2010rf} as well as their interplay, \eg the closeness of the respective pseudocritical regions,
properties of their critical points (such as critical exponents) \cite{Tiwari:2012yy}, 
the influence of external control parameters such as magnetic fields \cite{Ferrari:2012yw,Mizher:2010zb,Kamikado:2014bua},
or its interplay with hydrodynamical models \cite{Paech:2003fe,Herold:2013bi}.

For addressing the aforementioned question of phase structure imprints on the photon rate, the model has to be 
supplemented with an electromagnetic sector.
Following \cite{Mizher:2010zb,Kamikado:2014bua}, we do so by replacing the partial derivative $\partial^\mu$ by the $U(1)$-covariant 
derivative $D^\mu=\partial^\mu-ieQA^\mu$ ($e$ being the electromagnetic coupling strength and $Q$ the charge operator) 
and by adding the conventional
kinetic term for the gauge field (the photon field) $A^\mu$ with the field strength tensor 
$F^{\mu\nu} = \frac{i}{e}\Big[D^\mu,D^\nu\Big]$. 
This procedure corresponds to
adding the terms $\Lag_{k\gamma}$ and $\Lag_{e}$ to the Lagrangian \eqref{Lagrangian_01}:
\begin{align}
   \Lag_\eLSM     =& \Lag + \Lag_{k\gamma} + \Lag_e,\label{Lagrangian_05}\\
   \Lag_{k\gamma} =& -\frac14 F^{\mu\nu}F_{\mu\nu},\label{Lagrangian_06}\\
   \Lag_{e}       =& N_c\bar \psi eQ\slashed A \psi   +ieA^\mu \pi^-\partial_\mu\pi^+  
                                    -ieA_\mu \pi^+\partial^\mu\pi^-
                                    -e^2A_\mu A^\mu \pi^+\pi^- \label{Lagrangian_07}.
\end{align}
Having defined the model by $\Lag_\eLSM$ we go on by calculating the Euclidean generating functional $S_\eta$ 
for correlation functions.
We base our calculation on the path integral representation of $S_\eta$, 
but go beyond the standard
mean field approximation (MFA) and include lowest order fluctuations of the meson fields.
This seems necessary for the purpose of photon emission, since the electromagnetic field is expected to couple 
via derivatives to the charged pions, which are absent in the MFA approach.
\section{The generating functional}
\label{sec:GenFunc}
As mentioned above we go beyond MFA following the approximation scheme introduced in \cite{Mocsy:2004ab}.
For this purpose we have to calculate the photon emission in a consistent way. 
We achieve this goal by calculating the generating functional for Euclidean correlation functions to which we
can adopt - due to the formal similarity of the generating functional and the partition function - the 
approximations for the partition function made in \cite{Mocsy:2004ab}. By functional differentiation 
we then derive the photon propagator consistent with this approximation and apply the McLerran-Toimela formula
(see \eqref{McLerran_Toimela} below) to calculate the photon emission rate.

The path integral representation of the Euclidean generating functional for finite temperature and densities 
reads \cite{Peskin:1995ev,ZinnJustin:2002ru}
\begin{align}
   S_\eta \equiv&\ S[\SourcesSpqqA]\notag\\
          =& \int \Dqqsp\mathcal{D}[A] \label{GenFunc_01}\\
           & \times \exp\left\{-\int\limits_0^\beta d\tau \int\limits_{\mathbb{R}^3}d^3x \Lag_\eLSM(q,\bar q,\sigma,\pi)
           + \mu\bar\psi\gamma^0\psi
           + \bar q\eta_q + \bar\eta_q q + \eta_\sigma \sigma + \eta^\pi_a\pi^a + \eta^\gamma_\mu A^\mu\right\}, \notag
\end{align}
where $\beta$ is the inverse temperature $T^{-1}$ and $\SourcesSpqqA$ denote the sources of the respective fields.
(The measure $\mathcal{D}[A]$ refers to an path integral over gauge independent field configurations,
see \cite{Faddeev:1967fc}.)
The source term for the pions, $\vec \eta_\pi \vec \pi = \eta_\pi^1\pi_1 + \dots +\eta_\pi^3\pi_3$, can be rewritten
in terms of the charged pions according to $\vec \eta_\pi \vec \pi = \eta_\pi^-\pi^+ + \eta_\pi^+\pi^- + \eta_\pi^0\pi^0$
with $\chi^\pm = \sqrt{2}^{\ -1}(\chi^1 \mp i\chi^2)$, $\chi^0 = \chi^3$ and $\chi \in \{\pi,\eta_\pi\}$. 
\subsection{Integrating out the photons}
The path integral over the photon field configurations being quadratic in the fields 
(for the gauge fixing, the standard covariant choice 
$\Lag_\fix = \xi^{-1}(\partial A)^2$ is made) can be evaluated exactly leading to
\begin{align}
   S_\eta =&\int\Dqqsp \sqrt{\det \Gph{\mu\nu}}\notag\\
          & \times\exp\bigg\{\int dx^4 \big(\Lag- \mu \bar \psi \gamma^0\psi 
                                +\bar\eta_q \psi + \bar \psi \eta_q 
                                + \eta_\sigma \sigma + \vec \eta_\pi \vec \pi\big)\bigg\} \label{GenFunc_02}\\
          & \times\exp\bigg\{\int dz^4d{z'}^4(J_\gamma^\mu(z) + \eta^\mu_\gamma(z))\Gph[(z,z')]{\mu\nu}(J_\gamma^\nu(z')+\eta^\nu_\gamma(z'))\bigg\}\notag
\end{align}
with the electromagnetic current 
\begin{align}
   J_\gamma^\mu(z) &= -\bar \psi(z) e\hat Q\gamma^\mu \psi(z) 
                      - \pi^+(z) ie\partial^\mu \pi^-(z) 
                      + \pi^-(z) ie\partial^\mu \pi^+(z) 
\end{align}
and the perturbative photon propagator $\Gph{\mu\nu}$ formally defined by
\begin{align}
   \big(G^0_\gamma\big)^{-1}_{\mu\nu} &= \Big[g_{\mu\nu} \square - \left(1-\xi^{-1}\right)\partial_\mu\partial_\nu - e^2 \pi^+\pi^-g_{\mu\nu}\Big]\label{PhotonProp_01}.
\end{align}
\subsection{Integrating out the fermions}
The next step is to integrate out the quarks resulting in a fermion determinant, which is written as the exponential of a 
functional trace (\ie a momentum-integral of traces over internal (\ie Dirac- and flavor-) indices) 
and an exponential with source terms:
\begin{align}
   S_\eta =&\int\Dsp \sqrt{\det \Gph{\mu\nu}}\notag\\
           & \times\exp\bigg\{\int dx^4 \big(\Lkm - U - \left(\Tr \ln \Gqu\right)(x,x) 
                           + \eta_\sigma \sigma + \vec \eta_\pi \vec \pi\big)\bigg\} \label{GenFunc_03}\\
           & \times\exp\bigg\{\int dz^4d{z'}^4(J_\gamma^\mu(z)+\eta^\mu_\gamma(z)) \Gph[(z,z')]{\mu\nu}(J_\gamma^\nu(z')+\eta^\nu_\gamma(z'))
                           +\bar \eta_q(z) \Gqu[(z,z')]\eta_q(z')\bigg\},\notag
\end{align}
with the quark propagator defined in \eqref{Lagrangian_02}.
Up to now the evaluation is exact. But, since the remaining mesonic part is not at all a Gaussian integral, we are forced to
employ several approximations in order to proceed.
\subsection{Derivative expansion of the fermion trace}
The term $\Tr\ln\Gqu$ in \eqref{GenFunc_03} is expanded w.r.t. derivatives of the meson fields similar to \cite{Fraser:1984zb,Aitchison:1985pp} 
yielding (see \ref{apdx_deriv_exp} for details)
\begin{align}
   \Tr\ln \Gqu =& \Tr\ln \Big[i\slashed \partial - m_q(\sigma,\vec\pi)\Big] + \Ord{\partial\sigma,\partial\vec \pi}  \label{det_G_q_01}\\
               =& \frac{1}{3(2\pi)^3}\int dp^3 \frac{p^2}{E_q}(1 + n_F(E_q) + n_{\bar F}(E_q)) + \Ord{\partial\sigma,\partial\vec \pi},\\
         E_q^2 =& m_q^2 + p^2,\\
         m_q^2 =& g^2(\sigma^2+\vec\pi^2).\label{quarkmass_01}
\end{align}
Assuming slowly varying meson fields the terms containing meson derivatives can be regarded small and are thus excluded from 
further calculations.
\subsection{Quadratic approximation of the meson potential}
The generating functional in \eqref{GenFunc_03} can be regarded as the generating functional of a purely mesonic theory
with the potential $V$:
\begin{align}
   V(z)       =& U_\eff(\sigma(z),\pi(z)) 
                   - \int d{z'}^4J_\gamma^\mu(z) \Gph[(z,z')]{\mu\nu} J_\gamma^\nu(z'),\label{effective_Pot_02}\\
   U_\eff     =& U(\sigma,\vec\pi) - \Omega_\qq(\sigma,\vec\pi)\label{effective_Pot_01},\\
   \Omega_\qq =& \frac{2N_FN_c}{3(2\pi)^3}\int dp^3 \frac{p^2}{E_q}\Big(1 + n_F(E_q) + n_{\bar F}(E_q)\Big).\label{fermion_pot_01}
\end{align}
The effective potential $U_\eff$ is approximated by a quadratic potential $\bar U$ defined by the conditions
\begin{align}
   \langle\bar U\rangle =& \langle U_\eff \rangle , \label{Ueff_fix_01}\\
   \diff{\bar U}{\sigma,\pi}\Bigg|_{\genfrac{}{}{0pt}{1}{\sigma=\langle\sigma\rangle}{\vec\pi=0}} =& 0, &\text{ with $\langle\sigma\rangle$ determined by }& &
   \left\langle \diff{U_\eff(\langle\sigma\rangle+\Delta,\vec \pi)}{\Delta,\pi}\right\rangle=&0 \label{Ueff_fix_02}
\end{align}
and with $\langle f(\sigma,\vec\pi)\rangle$ being the ensemble average w.r.t. $\sigma$ and $\vec\pi$ configurations
according to the self consistently chosen probability density $\rho$ given below in \eqref{Prop_Distr_01} 
(\cf \eqref{average_01} below for the averaging).
The condition \eqref{Ueff_fix_01} fixes the zero-order coefficients in $\bar U$ and \eqref{Ueff_fix_02} the first order
coefficients. The non-vanishing second order coefficients (which we name $m_\sigma^2$ and $m_\pi^2$) 
have to be chosen according to  
\begin{align}
     \diff{^2 \bar U}{\sigma^2} \equiv m_\sigma^2 &= \left\langle \diff{^2 U_\eff}{\sigma^2} \right\rangle,
   & \diff{^2 \bar U}{\pi^2}    \equiv m_\pi^2    &= \left\langle \diff{^2 U_\eff}{\pi^2} \right\rangle \label{mass_def_01}
\end{align}
for being consistent to the respective propagator pole mass (see \eqref{meson_prop_01} below), 
calculated for the approximated theory with Lagrangian
\begin{align}
   \Lagbar&= \Lag_\text{km} - \bar U(\sigma,\pi).
\end{align}
The 2nd order mixed term in $U_\eff$ vanishes, since $U_\eff$ is an even function of $\vec\pi$ as the inspection of 
\eqref{Lagrangian_04}, \eqref{quarkmass_01} and \eqref{effective_Pot_01} reveals.
The thus defined approximate potential 
\begin{align}
   \bar U =&  \langle U_\eff \rangle + \frac12 m_\pi^2(\vec \pi^2-\langle \vec\pi^2\rangle) + \frac12 m_\sigma^2(\sigma^2 - \langle \sigma^2\rangle )\label{approx_pot_01}
\end{align}
induces via the accordingly approximated partition function $\bar Z$ 
a probability distribution $\rho$ for the meson fields 
(for further convenience we chose to shift the sigma field by its thermal expectation value, 
$\sigma = \Delta + \langle\sigma\rangle$)
\begin{align}
   \rho(\sigma,\vec\pi) =& {\bar Z}^{-1}\exp\left\{\int dx^4\Lag_{km}-\bar U\right\}\\
                        =& \frac{1}{\sqrt{2\pi \langle \Delta^2 \rangle}} 
                                  \exp\left\{\frac{-\Delta^2}{2\langle \Delta^2 \rangle}\right\}
                          \sqrt{\frac{2}{\pi}\left(\frac{3}{\langle \vec \pi^2 \rangle}\right)^3} 
                                  \exp\left\{\frac{-3\vec\pi^2}{2\langle \vec\pi^2 \rangle}\right\},\label{Prop_Distr_01}
\end{align}
where we already exploited the fact that all relevant functions are even functions of $\vec \pi$, 
leading to the following form of the variances ($\variance f = \langle f^2\rangle -\langle f\rangle^2$) for the meson fields
\begin{align}
   \langle \Delta^2 \rangle &= \variance \sigma =  \frac{1}{(2\pi)^3}\int d^3p\left(\frac{1}{2E_\sigma}+\frac{1}{E_\sigma}n_B(E_\sigma)\right),\label{fluct_si_01}\\
   \langle \vec\pi^2\rangle &= 3\,\variance \pi_i =  \frac{3}{(2\pi)^3}\int d^3p\left(\frac{1}{2E_\pi   }+\frac{1}{E_\pi   }n_B(E_\pi)\right),\label{fluct_pi_01}
\end{align}
with the dispersion relations $E_{\sigma,\pi}^2 = m_{\sigma,\pi}^2+ \vec p^2$. 
With \eqref{Prop_Distr_01} the ensemble averages $\langle f\rangle $ of a meson dependent function
can be calculated according to
\begin{align}
   \langle f(\sigma,\vec \pi)\rangle =& \int d\Delta \int d|\vec\pi| \rho(\Delta + \langle\sigma\rangle,|\vec\pi|) f(\Delta + \langle\sigma\rangle,|\vec\pi|),\label{average_01}
\end{align}
where we let $f$ only depend on $|\vec \pi|$, since that is the only case relevant for our calculation.
Equations \eqref{fluct_si_01} and \eqref{fluct_pi_01} represent consistency conditions for the choice of the second order
coefficients, \ie the meson mass parameters, as they can be calculated via the induced probability distribution \eqref{Prop_Distr_01}
as well as by differentiation of the thermodynamical potential (which is related to the generating functional, as discussed
below in \ref{apdx_ThDyn}). The equations \eqref{Ueff_fix_02}, \eqref{mass_def_01}, \eqref{fluct_si_01} and 
\eqref{fluct_pi_01} represent a set of five equations for $m_\pi$, $m_\sigma$, $\langle\sigma\rangle$, $\langle \Delta^2\rangle$ and 
$\langle \vec \pi^2\rangle$ which have to be solved simultaneously (\cf \cite{Mocsy:2004ab}, eqs. (51), (52), (17), (37) 
and (38) with (42) and (47) inserted into (17)).
The quark source term in \eqref{GenFunc_03} is treated by expanding $\Gqu$ w.r.t. the meson fields 
(\cf \mbox{\ref{apdx_invert_mtrx}}) and replacing the meson fields afterwards by the variation w.r.t. the
corresponding sources.
\subsection{Isolating the electromagnetic contribution}
We now want to treat the electromagnetic contribution in \eqref{effective_Pot_02} as a small perturbation to $S_\eta$.
Therefore, we first expand (\cf \ref{apdx_invert_mtrx}) the photon propagator $\Gph{\mu\nu}$ w.r.t. powers of 
$e^2$ and 
afterwards the exponential of the current term into a Taylor series:
\begin{align}
   \exp\bigg\{\int&dz^4d{z'}^4(J_\gamma^\mu(z)+\eta^\mu_\gamma(z)) \Gph[(z,z')]{\mu\nu}(J_\gamma^\nu(z')+\eta^\nu_\gamma(z')) \bigg\}\\
              = 1 & + \int dz^4d{z'}^4(J_\gamma^\mu(z)+\eta^\mu_\gamma(z)) \Big(\bar G^\gamma_{\mu\nu}(z,z')\notag\\
                  & + e^2\int dz''\bar G^\gamma_{\mu\rho}(z,z'')g^{\rho\kappa}\pi^+(z'')\pi^-(z'')\bar G^\gamma_{\kappa\nu}(z'',z') + \Ord{e^4}\Big)
                    (J_\gamma^\nu(z')+\eta^\nu_\gamma(z')) + \Ord{J_\gamma^4}\label{em_contrib_02}.
\end{align}
Since $J_\gamma^\mu = \Ord{e}$ the terms up to $\Ord{e^2}$ are
\begin{align}
   \exp\bigg\{\int&dz^4d{z'}^4(J_\gamma^\mu(z)+\eta^\mu_\gamma(z)) \Gph[(z,z')]{\mu\nu}(J_\gamma^\nu(z')+\eta^\nu_\gamma(z')) \bigg\}\\
              = 1 & + \int dz^4d{z'}^4(J_\gamma^\mu(z)+\eta^\mu_\gamma(z)) \bar G^\gamma_{\mu\nu}(z,z')
                   (J_\gamma^\nu(z')+\eta^\nu_\gamma(z')) + \Ord{J_\gamma^4}\notag\\
                  & +\int dz^4d{z'}^4\int d{z''}^4 \eta^\mu_\gamma(z) e^2\bar G^\gamma_{\mu\rho}(z,z'')g^{\rho\kappa}\pi^+(z'')\pi^-(z'')\eta^\nu_\gamma(z)\bar G^\gamma_{\kappa\nu}(z'',z') + \Ord{e^3}.\label{em_contrib_01}
\end{align}
Finally, $\pi^+(z)\pi^-(z)$ is replaced by $\delta/\delta \eta_\pi^-(z)\delta/\delta \eta_\pi^+(z)$ and 
$J$ by $\dJem{}$ as defined by 
\begin{align}
    \dJem[(z)]{\mu} &= -\variation{}{\eta_q(z)} e\hat Q\gamma^\mu \variation{}{\bar \eta_q(z)} 
                      - \variation{}{\eta_\pi^-(z)} ie\partial^\mu \variation{}{\eta_\pi^+(z)} 
                      + \variation{}{\eta_\pi^+(z)} ie\partial^\mu \variation{}{\eta_\pi^-(z)}.
\end{align}
With this replacements, \eqref{em_contrib_01}
can be pulled out of the path integral.
\subsection{Integrating out the meson fields}
The remaining integrand of the path integrals in \eqref{GenFunc_03} is Gaussian yielding
\begin{align}
   S_\eta =&\sqrt{\det \Gph{\mu\nu}}\sqrt{\det \Gpi}^3\sqrt{\det \Gsi}\exp\left\{-\int d^4x\langle U_\eff\rangle
           +\frac12 m_\pi^2\langle\vec\pi^2\rangle +\frac12m_\sigma\langle\Delta^2\rangle\right\}\notag\\
          & \times\exp\bigg\{\int dz^4d{z'}^4\left(\dJem[(z)]{\mu} + \eta^\mu_\gamma(z) \right) 
                   \dGph[(z,z')]{\mu\nu}\left(\dJem[(z')]{\nu} + \eta^\nu_\gamma(z')\right)\bigg\}\label{GenFunc_07}\\
          & \times\exp\bigg\{\bar \eta_q(z) \dGqu[(z,z')]\eta_q(z') + \eta_\sigma(z)\Gsi(z,z') \eta_\sigma(z') 
           + \eta_\pi^+(z)\Gpi(z,z')\eta_\pi^-(z')+ \eta_\pi^0(z)\Gpi(z,z')\eta_\pi^0(z')
                            \bigg\}\notag,  
\end{align}
with $\dGqu$ and $\dGph{\mu\nu}$ obtained from $\Gqu$ and $\Gph{\mu\nu}$ by replacing 
$\sigma\rightarrow \delta/\delta \eta_\sigma$ and $\pi^a\rightarrow\delta/\delta \eta_{\pi^{-a}}$. 
The momentum space meson propagators can be found by an explicit evaluation of the Gaussian meson path 
integrals yielding 
\begin{align}
   G_\pi^{a\bar b}(p) =& \frac{\delta^{ab}}{p^2 -m_\pi^2},             &   
   \Gsi(p) =& \frac{1}{p^2 -m_\sigma^2}\label{meson_prop_01}   
\end{align}
with $a,b\in\{0,+,-\}$ denoting the charge of the respective pions, 
$\bar b \equiv -b$ and the mass parameters according to \eqref{mass_def_01}.
\section{The imaginary part of the photon propagator}
\label{sec:PhotonProp}
Having derived the formal basics we turn now to the derivation of photon emission rates. The wanted key
quantity is the imaginary part of the retarded photon propagator which allows to cast the rates into 
a kinetic theory formula.
\subsection{The photon propagator}
\label{sec:photon_prop}
\begin{figure}
   \centering
   \includegraphics[width=0.9\textwidth]{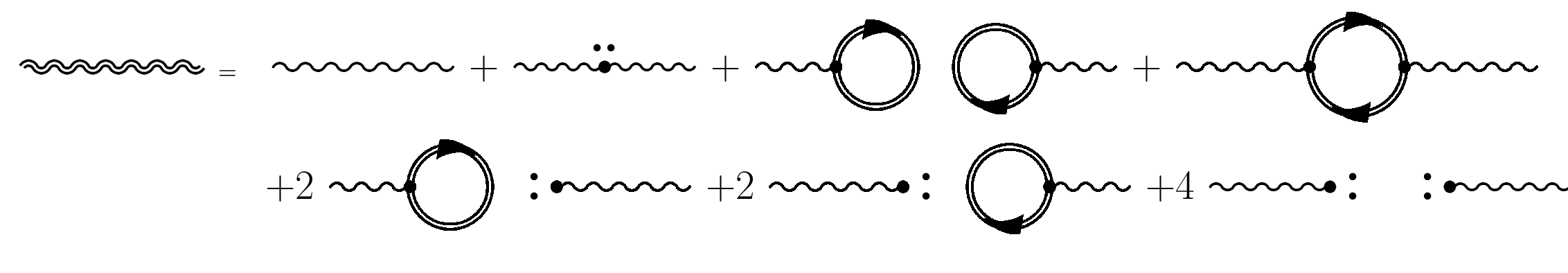}\\
   \includegraphics[width=0.9\textwidth]{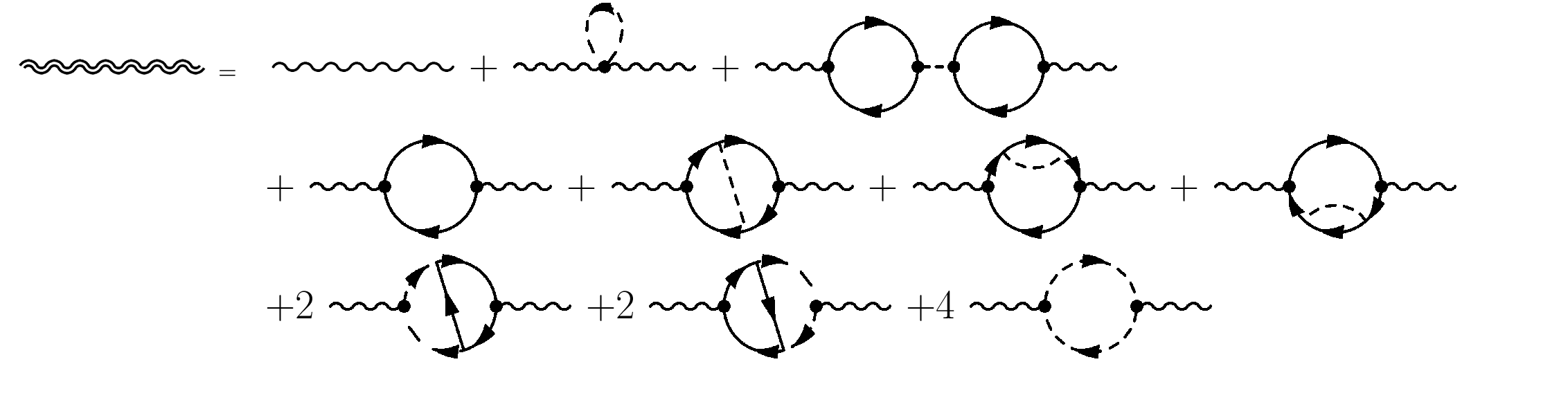}
   \caption{Upper panel: Diagrammatic representation of the electromagnetic current term in \eqref{Photon_Prop_03} 
   after applying the fermion
   source derivatives and setting the fermion sources zero.
   Lower panel: Diagrammatic representation of the electromagnetic current term after additionally 
   executing all meson source derivatives, both written down explicitly as dots next to a vertex
   and implicitly contained in the summed fermion propagator.
   Solid double lines represent the summed fermion propagator $\dGqu$, the double wavy line stands for the full
   photon propagator up to $\Ord{e^2}$ and $\Ord{g^2}$, the single solid lines represents $\Gquvev{}$, single wavy lines
   the perturbative photon propagator $\bar G^\gamma_{\mu\nu}$, dashed lines represent a sum over all meson field
   propagators $\Gpi$ and $\Gsi$ connected
   to the fitting vertices (dots). Arrows on dashed lines denote the direction of charge flow. (They appear only at lines
   connected to a photon vertex, so that only charged pions contribute to the diagram, for which the direction of 
   charge flow is well defined.)}
   \label{fig:photon_prop}
\end{figure}

The full photon propagator (within the above approximations), $\mathcal{G}$, can be calculated by varying $S_\eta$ w.r.t. the photon sources:
\begin{align}
   \mathcal{G}^\gamma_{\mu\nu}(x,y) = \frac{1}{S}\frac{\delta^2}{\delta\eta_\gamma^\mu(x)\delta\eta_\gamma^\nu(y)}S_\eta.
\end{align}
Executing the variations yields
\begin{align}
   \mathcal{G}^\gamma_{\mu\nu}(x,y) 
       =&   \bar G^\gamma_{\mu\nu}(x,y) 
          + e^2\frac{1}{S}\int dz\bar G^\gamma_{\mu\alpha}(x,z)\variation{}{\eta_\pi^-(z)}\variation{}{\eta_\pi^+(z)}g^{\alpha\beta}\bar G^\gamma_{\beta\nu}(z,y)S_\eta\Bigg|_{\eta=0}\notag\\
        & + \frac{1}{S}\int dz\int dz' \bar G^\gamma_{\mu\alpha}(x,z)\dJem[(z)]{\alpha}\dJem[(z')]{\beta}\bar G^\gamma_{\beta\nu}(z',y) S_\eta\Bigg|_{\eta=0},\label{Photon_Prop_03}\\
   \big(\bar G^\gamma_{\mu\nu}\big)^{-1} =& \Big[g_{\mu\nu} \square - \left(1-\xi^{-1}\right)\partial_\mu\partial_\nu\Big] = \Gph{\mu\nu}^{-1}+ e^2 \pi^+\pi^-g_{\mu\nu}.
\end{align}
First we execute the variations w.r.t. the fermionic sources. The result can be represented diagrammatically as exhibited 
in Fig.~\ref{fig:photon_prop} (upper panel). 
Looking at the upper panel of Fig.~\ref{fig:photon_prop} we want to stress that it represents an intermediate step 
of the calculation and the (conventional) prescription that only connected diagrams contribute cannot be applied yet. 
Comparing the upper panel of Fig.~\ref{fig:photon_prop} with \eqref{Photon_Prop_03} one can identify the 
first two diagrams with the first two terms in \eqref{Photon_Prop_03}. 
The rest of the diagrams corresponds to the $\Ord{e^2}$ contributions of the second line of \eqref{Photon_Prop_03} 
with the derivatives w.r.t. the fermionic sources carried out. 
Since the fermionic sources appear only as $\exp\big\{\int\int dz^4d{z'}^4\bar \eta_q(z) \dGqu[(z,z')]\eta_q(z')\big\}$ 
in $S_\eta$, every pair of (functional) derivatives w.r.t. fermionic sources corresponds 
to a quark propagator $\dGqu$ connecting the space-time arguments of the source derivatives, 
exactly as it is conventionally done, \eg in Feynman diagram calculus of QED correlation functions 
or self energy contributions. (However, since $\dGqu$ explicitly depends on meson source derivatives, 
disconnected fermionic loops can be connected in the next step by meson propagators 
leading to the lower panel of Fig.~\ref{fig:photon_prop}.)
The fermion propagators represented by double lines
are expanded according to \eqref{quark_prop_04}.
This expansion follows from the decomposition of $\Gqu$ into a part independent of the dynamical fields 
$\Delta$ and $\pi$ (but dependent on the average $\sigma$ field) and an interaction part that depends on 
$\Delta$ and $\pi$ according to
\begin{align}
   \left(\Gqu\right)^{-1}    =& \left(\Gquvev[]\right)^{-1} - g(\Delta + i\gamma^5\vec\tau \vec\pi),\label{quark_prop_01}\\
   \left(\Gquvev\right)^{-1} =& i\slashed \partial - g\langle\sigma\rangle.\label{quark_prop_03}
\end{align}
After doing all meson variations, symbolically denoted by dots next to the vertices in the upper panel of
Fig.~\ref{fig:photon_prop}, every pair of derivatives w.r.t. meson sources reduces to a propagator of the respective
meson (provided both variations are w.r.t. the same fields source). The resulting expression 
can be represented diagrammatically according to the lower panel of Fig.~\ref{fig:photon_prop}.
There, each solid line represents $\Gquvev$, each dashed line stands for $\Gsi{}+\Gpi{}$, each
wiggly line refers to $\bar G^\gamma_{\mu\nu}$ and each dot means the corresponding vertex factor.
\subsection{Determining the imaginary part of the photon propagator}
\begin{figure}
   \centering
   \subfigure{\includegraphics[width=0.25\textwidth,clip]{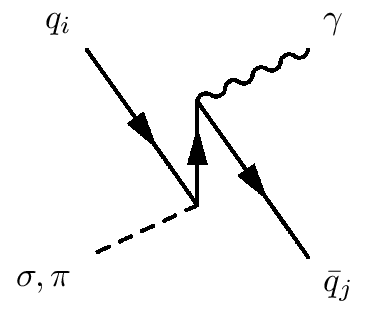}}
   \subfigure{\includegraphics[width=0.25\textwidth,clip]{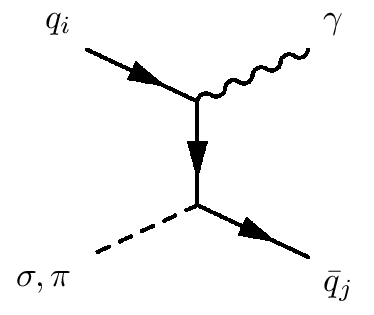}}
   \subfigure{\includegraphics[width=0.25\textwidth,clip]{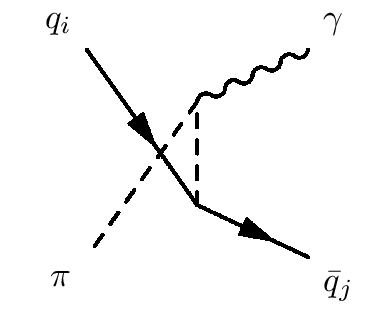}}\\
   \subfigure{\includegraphics[width=0.25\textwidth,clip]{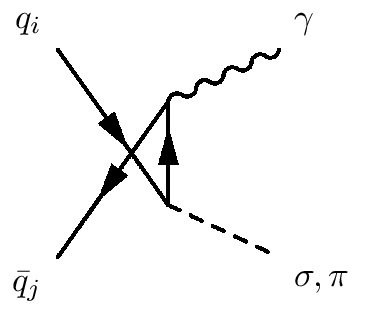}}
   \subfigure{\includegraphics[width=0.25\textwidth,clip]{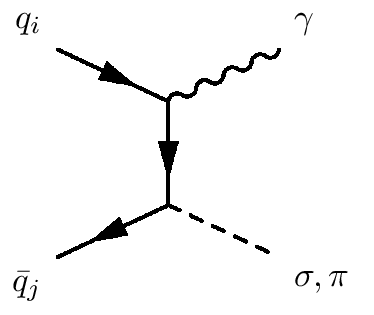}}
   \subfigure{\includegraphics[width=0.25\textwidth,clip]{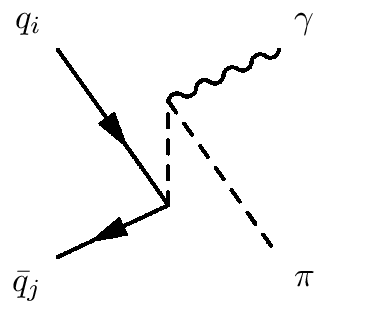}}
   \caption{Tree-level diagrams of the reactions \eqref{Compton_01}-\eqref{Annihil_01}. 
            Top row: diagrams for Compton scattering off quarks; 
            bottom row: annihilations.
            The matrix elements for Compton scattering off antiquarks can be obtained by inverting the 
            direction of the fermion arrows. Note that in the third column there is no $\sigma$ meson, since it is 
            uncharged. }
   \label{fig:matrix_elements}
\end{figure}

The propagators $\Gsi$, $\Gpi$ (see \eqref{meson_prop_01}) and $\Gquvev{}$ (see \eqref{quark_prop_03}) 
have the form discussed in \cite{Weldon:1983jn}. For the imaginary part of the diagrams in 
Fig.~\ref{fig:photon_prop}, one therefore has to cut through each diagram in any possible way that separates
the two vertices connected to external photon lines. 
Such a procedure leads to sets of (simpler) diagrams corresponding to processes of the type
$\phi_1,\dots,\phi_a \rightarrow \Phi_1,\dots,\Phi_b + \gamma$ with $a$ incoming and 
$b+1$ outgoing field quanta, one of which is a photon.
Denoting the diagrams in Fig.~\ref{fig:photon_prop} by $\mathcal{M}_{\gamma\rightarrow\gamma}^{(j)}$ and the 
diagrams obtained from cutting these by $\mathcal{M}^{(j,l)}_{\phi_1+\dots+\phi_a\rightarrow\Phi_1+\dots+\Phi_b+\gamma}$
one arrives for 
$\Im\mathcal{G}_{\gamma,\ret}^{\mu\nu} \sim \sum_j \Im\mathcal{M}_{\gamma\rightarrow\gamma}^{(j)}$ at
\begin{align}
   \Im \mathcal{G}_{\gamma,\ret}^{\mu\nu} =& 2\sum_{a,b,j,l}\int d\Omega_{ab} 
                         |\mathcal{M}^{(j,l)}_{\phi_1+\dots+\phi_a\rightarrow\Phi_1+\dots+\Phi_b+\gamma}|^2
                         n^{(i,1)}\cdots n^{(i,a)}(1\pm n^{(o,1)})\cdots(1\pm n^{(o,b)})(e^{\omega/T}-1),\label{Weldonformel_01}\\
    \int d\Omega_{ab}   =& \int \frac{d^3p_1}{2E^{(1)}_p(2\pi)^3}\cdots\frac{d^3p_a}{2E^{(a)}_p(2\pi)^3}
                               \frac{d^3q_1}{2E^{(1)}_q(2\pi)^3}\cdots\frac{d^3q_b}{2E^{(b)}_q(2\pi)^3}(2\pi)^4
                               \delta\left(k-\sum_{c} p^c +\sum_{d}q^d\right),
\end{align}
with $n^{(i/o,l)}$ being Fermi or Bose distribution functions  
(depending on the spin of the particle $l$ in the in ($i$) or out ($o$) state).
The summands in \eqref{Weldonformel_01} can be sorted w.r.t. the number $=a+b+1$ of participating fields. The inspection 
of the phase space regions over which one has to integrate on the rhs. of \eqref{Weldonformel_01} yields zero for all
summands with $a+b\leq2$ since the phase space vanishes in these cases, at least if all field quanta - except the
photons - are massive (as it is the case in our model). The first non zero terms have $a+b=3$ and correspond to the
$2\rightarrow2$ processes
\begin{align}
   q_i + \sigma,\pi      &\rightarrow q_j + \gamma        & \hspace{-1em}\text{(Compton scatterings off quarks)},\label{Compton_01}\\
   \bar q_i + \sigma,\pi &\rightarrow \bar q_j + \gamma   & \hspace{-1em}\text{(Compton scatterings off antiquarks)},\label{ACompton_01}\\
   q_i + \bar q_j        &\rightarrow \sigma,\pi + \gamma & \hspace{-1em}\text{(annihilations).}\label{Annihil_01}
\end{align}
The corresponding diagrams are depicted in Fig.~\ref{fig:matrix_elements}, and the polarization, spin and flavor 
summed/averaged squared matrix elements are listed in \ref{apdx_mtrx_elem}.
\subsection{Photon emission rate}
\label{sec:Rates_deriv}
From the imaginary part of the photon propagator, the photon rate can be determined according to the 
McLerran-Toimela formula \cite{McLerran:1984ay}
\begin{align}
   \omega\frac{d^7 N}{\dkdx} &= \frac{g_{\mu\nu}}{(2\pi)^3}\Im G_{\gamma,\ret}^{\mu\nu}(k^2=0, \omega) n_B(\omega). \label{McLerran_Toimela}
\end{align}
With \eqref{Weldonformel_01} we get as the leading terms in the above mentioned expansion w.r.t. the number of participating
particles
\begin{align}
   \omega\frac{d^7 N}{\dkdx} =& 2\frac{g_{\mu\nu}}{(2\pi)^{12}}\sum_{i,j}
   \int \frac{d^3p_1}{2E^{(1)}_p}\frac{d^3p_2}{2E^{(2)}_p} \frac{d^3q}{2E_q}(2\pi)^4
                              \delta(k-p^1-p^2 +q)|\mathcal{M}^{(i,j)}_{\phi_1+\phi_2\rightarrow\Phi_3+\gamma}|^2
                         n^{(1)}n^{(2)}(1\pm n^{(3)})\label{kinetic_theory_01}
\end{align}
which resembles the formula for the production of photons in $2\rightarrow2$ processes within a kinetic theory approach.

One might ask the question of what we have achieved with the above derivation that goes beyond the mere use of formula 
\eqref{kinetic_theory_01} which could naively be used right from the beginning. The very necessity to 
derive \eqref{kinetic_theory_01} is to identify precisely the mass parameters $m_\sigma, m_\pi, m_q$ to be used for the 
calculation of the photon rates. The meson masses have to be chosen to be the second order coefficients of the 
approximative potential $\bar U$ according to \eqref{mass_def_01}. However, the correct fermion mass is less obvious.
Reference \cite{Mocsy:2004ab} suggests to use $M_n=(\langle m_q(\sigma,\vec \pi)^n\rangle)^{1/n}$ for conveniently
chosen values $n$ without convincing justification for this choice. As the resulting quark mass 
does not strongly depend on $n$, we  
used $n=2$ in a previous work \cite{Wunderlich:2015rwa} on the subject. With the derivation above this issue can 
be settled by choosing
\begin{align}
   m_q = g\langle\sigma\rangle \label{mass_def_02}
\end{align}
in order to arrive at a representation of the photon propagator in terms of Feynman
diagrams which can simply be cut to obtain the imaginary part of each diagram leading to the kinetic-theory-like formula 
\eqref{kinetic_theory_01}. In other words, one may say that the
$M_n$ can be regarded as ``thermodynamical mass'' parameters since they are reasonable choices to be used in thermodynamic integrals,
but fail to be used in perturbative calculations as mass parameters for the quark propagators.
Although the choice of $n$ does affect $M_n$ only sightly, there is a large difference between $M_n$ and $m_q$ in
the chirally restored phase. (In the chirally broken phase the meson field fluctuations are smaller which brings $M_n$ and
$m_q$ closer together.).
Thus the validity of 
\eqref{kinetic_theory_01} relies on the consistent choice \eqref{mass_def_02} for the mass parameter
to be used in the fermion propagators.

\section{Evaluation of photon rates}
\label{sec:Rates_eval}
\subsection{Parameter fixing}
To be explicit one has to fix the parameters of the Lagrangian defined by \eqref{Lagrangian_01}-\eqref{Lagrangian_04}. 
The simplest way to do so is to set the vacuum masses
of the fields as well as the vacuum expectation value of the $\sigma$ field to specific values.
In mean field approximation and the linearized fluctuation approximation without vacuum fluctuations the relations 
between the parameters and the vacuum field properties can be given by the set of equations
\begin{align}
   (\mSiVac)^2-(\mPiVac)^2 =& 2\lambda \vSiVac^2, & (\mSiVac)^2 - 3(\mPiVac)^2 =& 2\lambda\zeta,\\
   (\mPiVac)^2\vSiVac      =& H,                  & \mNuVac =& 3g\vSiVac,\label{param_def_01}
\end{align}
with the nucleon mass at $T=\mu=0$ taken as $\mNuVac=3m_q^\text{vac}$. Typically, one chooses for $\mSiVac$, $\mPiVac$ and
$\mNuVac$ the PDG values \cite{Agashe:2014kda} and $\vSiVac=f_\pi$. However, this is not
strictly required for making contact to QCD, since at low temperatures many other degrees of freedom are
relevant, which could easily shift these values one way or the other thus giving some flexibility to the parameters
as long as typical mass scales are kept at the order of $\Lambda_{QCD}$.
Throughout this paper we use the values of $\mSiVac,\mPiVac,\mNuVac$ and $\vSiVac$ to identify the parameter sets,
which are collected in Tab.~\ref{tab:parameters}.
\begin{table}
   \centering
   \footnotesize{
   \begin{tabular}{c}
      \hline\hline
      \begin{tabular}{c|c@{\hspace{1em}}c@{\hspace{1em}}c@{\hspace{1em}}cp{0.5em}c@{\hspace{1em}}c@{\hspace{1em}}c@{\hspace{1em}}c}
      \setlength{\tabcolsep}{3em}
      &&&&&&&&\\[-1em]
      [$\MeV$]& $\mNuVac$ & $\mSiVac$ & $\mPiVac$ & $\vSiVac$ && 
              $T_{c}(\mu=0)$  & $\mu_c(T=0)$ & $T_\CEP$ & $\mu_\CEP$\\\cline{1-5}\cline{7-10}
          A & 936.0  &  700.0 & 138.0 & 92.4  && 148.3 & 328   & 72.5   & 279.5 $\vphantom{\sum\limits^1}$\\
          B & 1170.0 & 1284.4 & 138.0 & 90.0  && 194.6 & 430   & 97.0   & 377.5 \\
          C & 1080.0 &  700.0 & 138.0 & 90.0  && 140.3 & 324   & 98.0   & 216.0 \\
      \end{tabular}
      \\\hline\hline
   \end{tabular}
   }
   \caption{
            Parameter sets used for the analysis. The parameters $\mNuVac$, $\mSiVac$, $\mPiVac$, $\vSiVac$ can be
            mapped to the parameters $g$, $\lambda$, $\zeta$, $H$ of the Lagrangian \eqref{Lagrangian_01} 
            by \eqref{param_def_01}.
            These parameters yield the cross over temperature $T_{c}(\mu=0)$ at vanishing
            chemical potential, 
            the critical chemical potential at zero temperature $\mu_c(T=0)$ and the coordinates for the $\CEP$
            ($T_\CEP, \mu_\CEP$) 
            given in the last columns. All quantities are in units of $\MeV$ as indicated in the upper left corner of the table.
            }
   \label{tab:parameters}
\end{table}
\subsection{Differential Spectra}
\label{sec:diff_spectra}
\begin{figure}
   \centering
   {\includegraphics[trim=0mm 3mm 5mm 18mm,width=0.48\textwidth,clip]{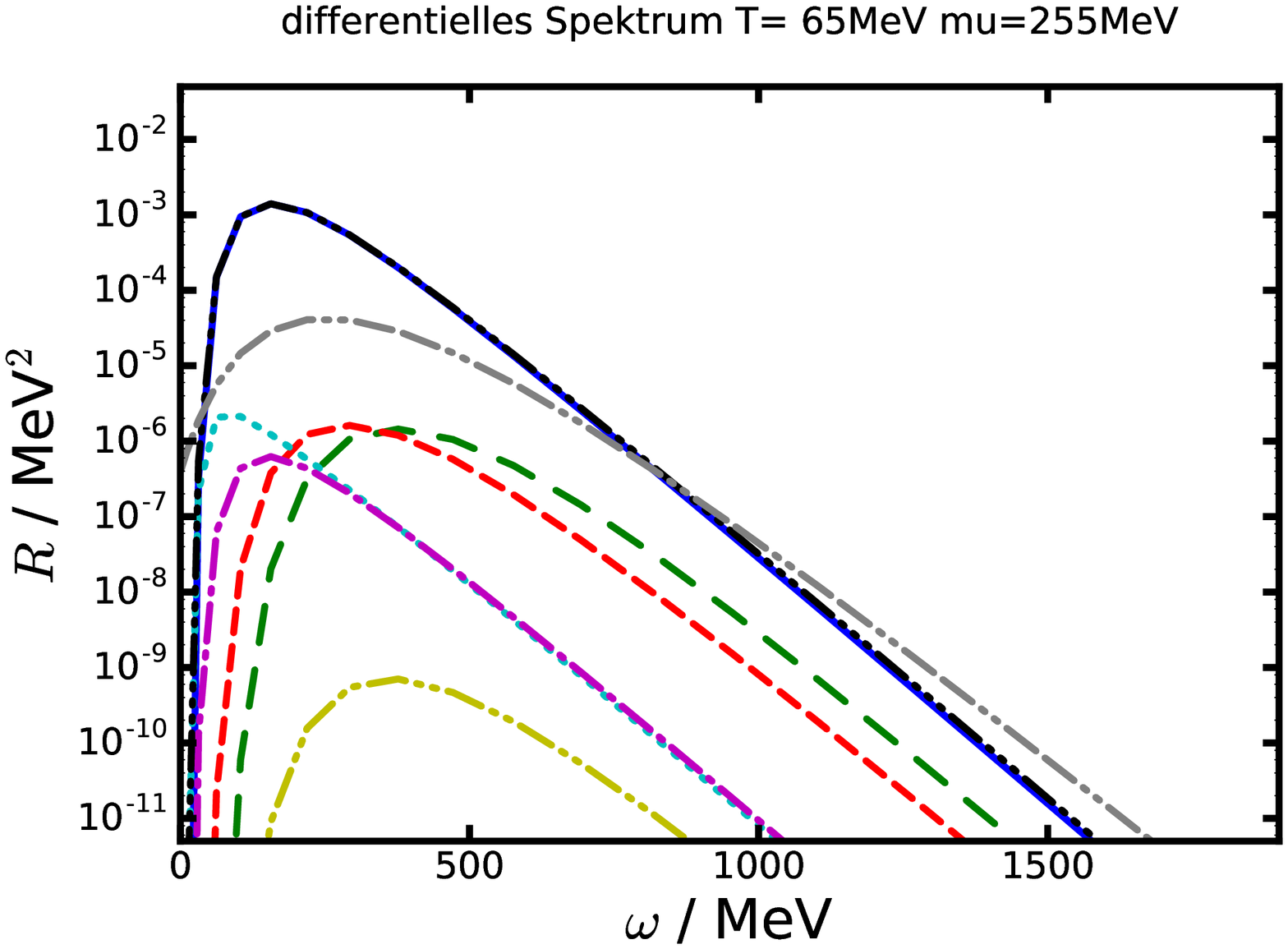}}
   {\includegraphics[trim=0mm 3mm 5mm 18mm,width=0.48\textwidth,clip]{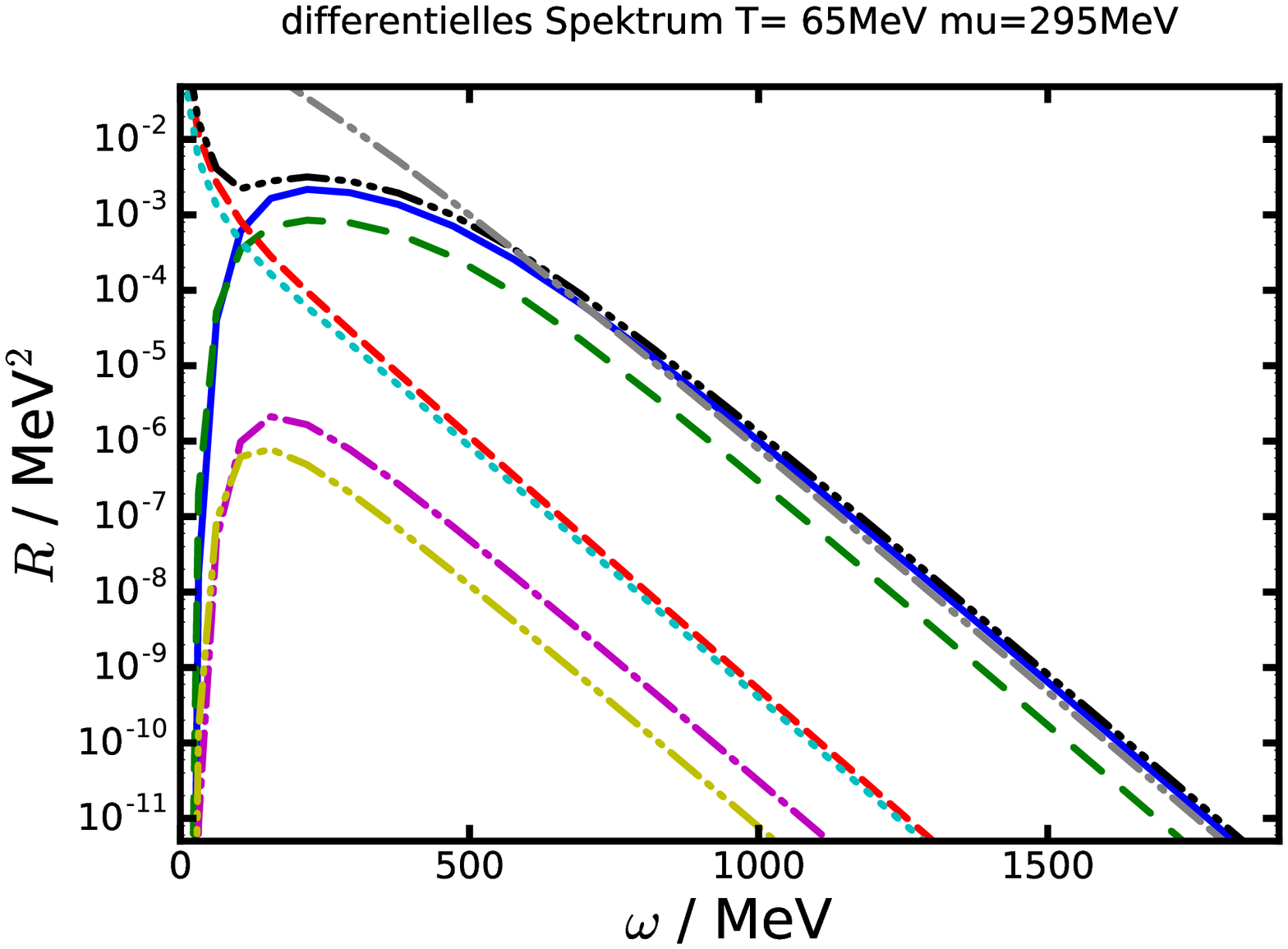}}
   \caption{Differential photon spectra $R=\omega d^7 N / \dkdx$ as functions of photon energy $\omega$ 
   for parameter set A. 
   Left panel: chirally broken phase at 
   $T=65\MeV$ and $\mu=255\MeV$; 
   right panel: restored phase at $T=65\MeV$ and $\mu=295\MeV$.
   The curves correspond to the partial rates for the processes 
   $\bar q + \sigma \rightarrow \bar q + \gamma$ (dash-double-dotted yellow curve), 
   $\bar q + \pi \rightarrow \bar q + \gamma$ (dash-dotted violet curve),
   $q + \bar q\rightarrow \sigma + \gamma$ (dotted light blue curve), 
   $q + \bar q\rightarrow \pi + \gamma$ (short dashed red curve),
   $q + \sigma \rightarrow q + \gamma$ (long dashed green curve), 
   $q + \pi    \rightarrow q + \gamma$ (solid dark blue curve)
   and to their sum (dash-triple dotted black curve), respectively.
   The gray double-dash-double-dotted curve 
   in the left panel is calculated with the parametrization 
   given in \cite{Heffernan:2014mla} and in the right panel it
   is the AMY rate \cite{Arnold:2001ms} for which the strong coupling $g_s$ was set 
   to the value of the quark-meson coupling $g$ of parameter set A.
   Since the masses $m_{\sigma,\pi,q}$ play an essential role for the photon emissivity
   we quote their values and supply the number densities in the heat bath 
   rest frame:
   }
     \footnotesize
     \begin{tabular}{l|ccc|ccc}
       \hline\hline
       \vphantom{${T^3}^3$} 
                   & $m_\sigma/\MeV$ & $m_\pi/\MeV$ & $m_q/\MeV$ & $n_\sigma/\fm^{-3}$ & $n_\pi/\fm^{-3}$    & $n_q/\fm^{-3}$\\
       \hline\vphantom{${f^3}^3$}
       left panel  & 504             & 151          & 282        & $2.65\times10^{-5}$ & $4.73\times10^{-3}$ & 0.205\\
       right panel & 283             & 333          &  55        & $3.93\times10^{-4}$ & $6.61\times10^{-4}$ & 0.971\\
       \hline\hline
    \end{tabular}
    \normalsize
   \label{fig:diff_spectra}
\end{figure}

In Fig.~\ref{fig:diff_spectra}, the differential photon spectra $\omega d^7 N / \dkdx$ are 
depicted for the individual channels \eqref{Compton_01}-\eqref{Annihil_01} as well as for their sum. 
A first inspection shows several aspects: (i) For large photon energies, $\omega\gtrsim 600\MeV$, all 
partial rates decrease exponentially $\propto \exp\{-\omega/T\}$. (ii) In the chirally
restored phase (see right panel of Fig.~\ref{fig:diff_spectra}), 
the partial rates for processes involving different mesons are either approximately degenerate (for the annihilations)
or differ only by a factor of about three.
This is a manifestation of the chiral symmetry restoration which leads to approximately degenerate meson masses, which 
in turn lead to
similar kinematics for the processes involving chiral partners. The difference between pion-involving 
and sigma-involving partial rates can be atributed to the different multiplicities and the charge carried by two of the 
pions. Conversely, in the chirally broken phase no such striking similarity
can be observed (see left panel of Fig.~\ref{fig:diff_spectra}). (iii) In the chirally restored phase, at 
$\omega \gtrsim 200\MeV$ there is a clear hierarchy of the rates from different processes:
The partial rate from Compton processes with quarks is much larger than the partial rate from annihilations which is also
much larger than the rate from Compton processes with anti-quarks. As pointed out in \cite{Wunderlich:2015rwa}
the reason for this hierarchy is the exponential suppression of incoming anti-particles at finite chemical potential, leading
to a suppression $\propto \exp\{-\mu/T\}$ of the annihilation processes and a suppression $\propto\exp\{-2\mu/T\}$ for the 
anti-Compton processes w.r.t. the partial rates from Compton processes with quarks. (iv) In the chirally restored
phase the annihilation rates diverge at $\omega\rightarrow 0$, see right panel of Fig.~\ref{fig:diff_spectra}. This is caused by  infrared  divergencies 
of the matrix elements
which are exponentially suppressed $\propto\exp\{-1/\omega\}$ if the sum of the incoming masses is larger than the mass
of the outgoing particle that is not a photon. Such an inequality holds necessarily for all Compton and anti-Compton
processes but may be violated for the annihilation processes; \cf \cite{Wunderlich:2015rwa} for a further discussion
of that issue. For a comparison with rate calculations in the literature we exhibit
in the left panel of Fig.~\ref{fig:diff_spectra}
the rate calculated with the parametrization given in \cite{Heffernan:2014mla} for the emission from the hadronic phase 
and in the right panel the AMY rate \cite{Arnold:2001ms} 
for the deconfined phase, however, for keeping the comparison as simple as possible both at $\mu=0$.(For the strong
coupling $g_s$ in the AMY rate we use the value of the quark-meson coupling $g$, corresponding to $\alpha_s=0.91$
for parameter set A.)
\subsection{Photon rate over the phase diagram}
\begin{figure}
   \centering
   {\includegraphics[trim=6mm 16mm 14mm 34mm,width=0.49\textwidth,clip]{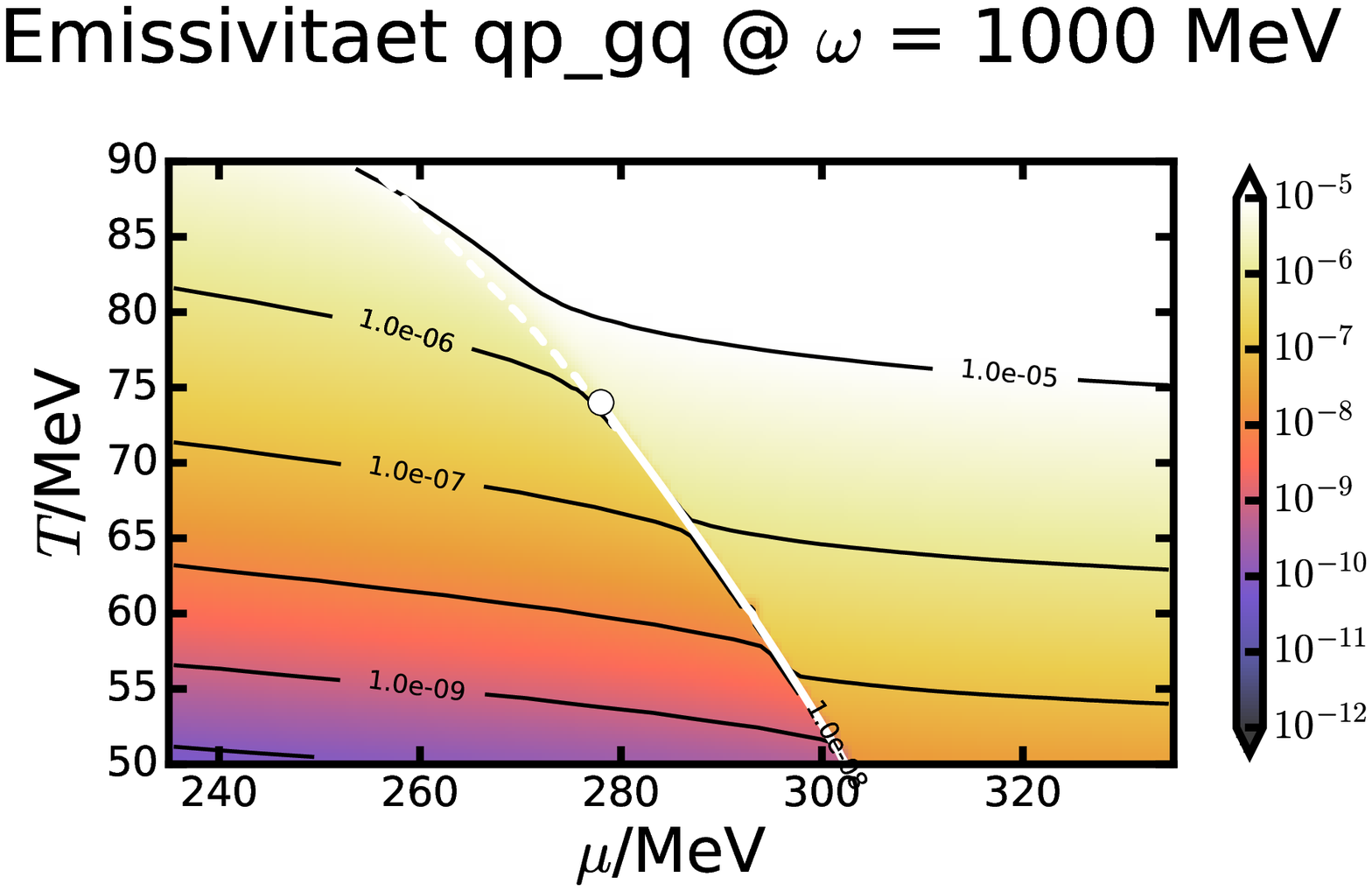}}
   {\includegraphics[trim=0mm 2mm 5mm 11mm,width=0.47\textwidth,clip]{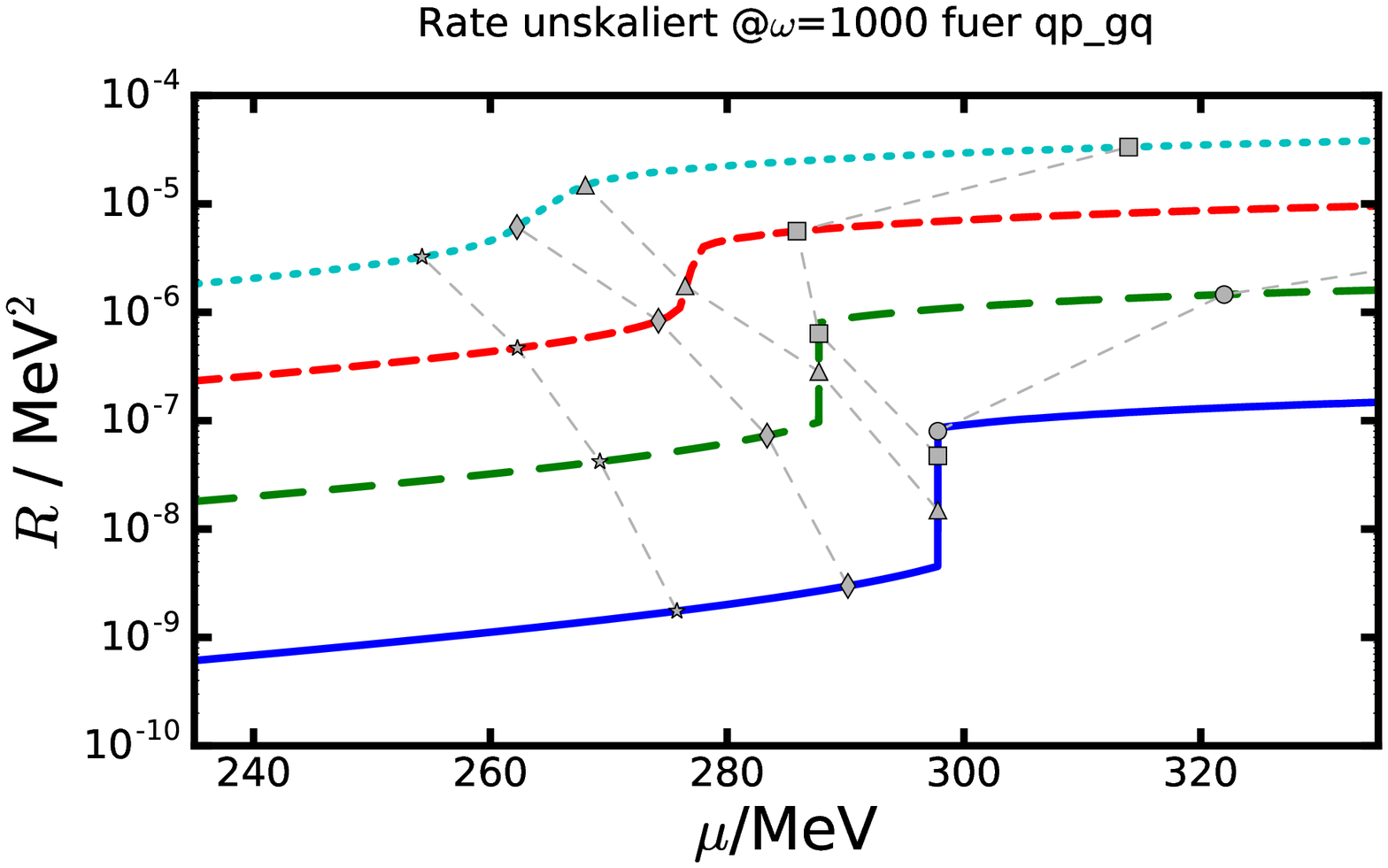}}\\
   {\includegraphics[trim=6mm 16mm 14mm 34mm,width=0.49\textwidth,clip]{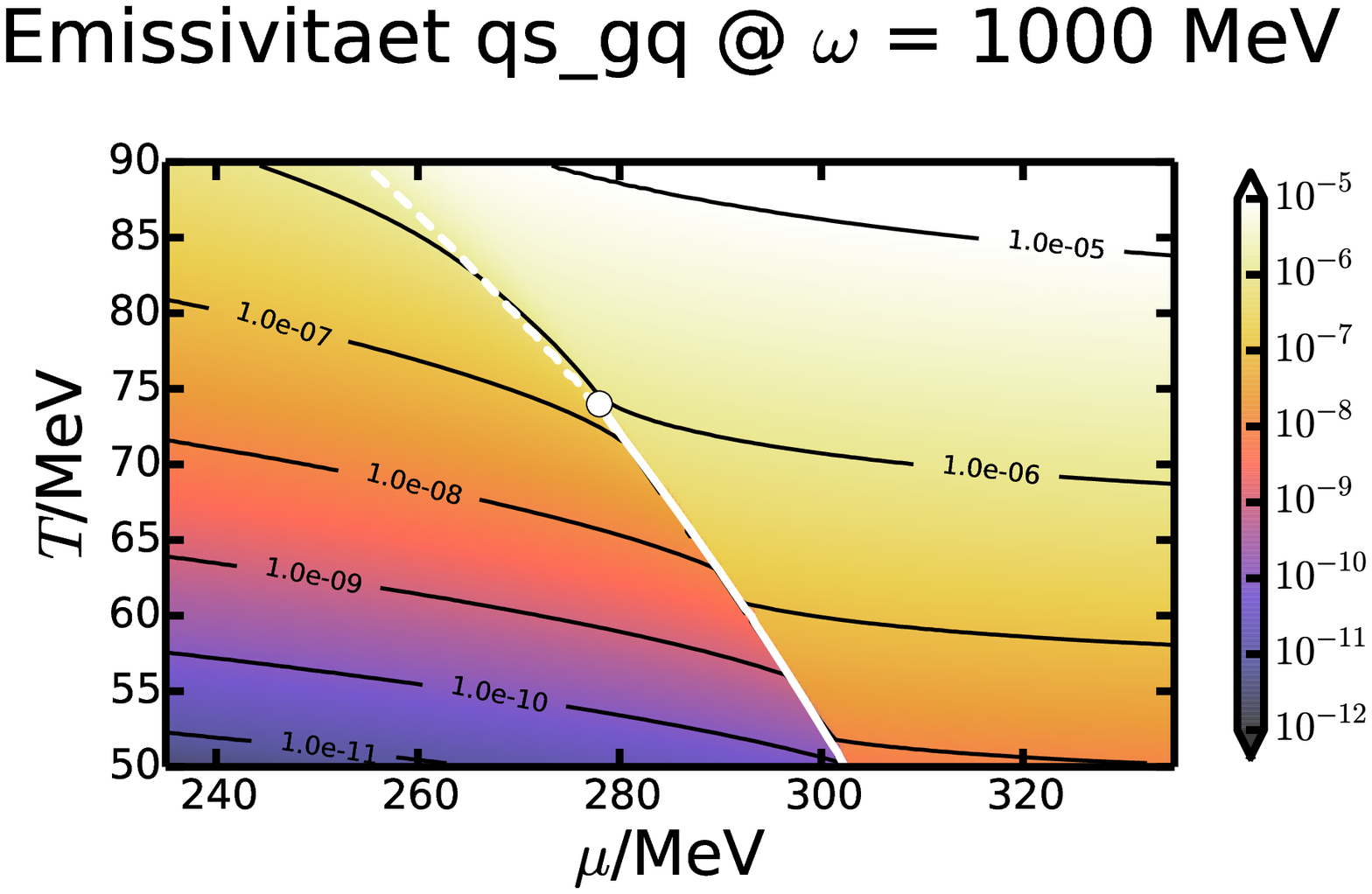}}
   {\includegraphics[trim=0mm 2mm 5mm 11mm,width=0.47\textwidth,clip]{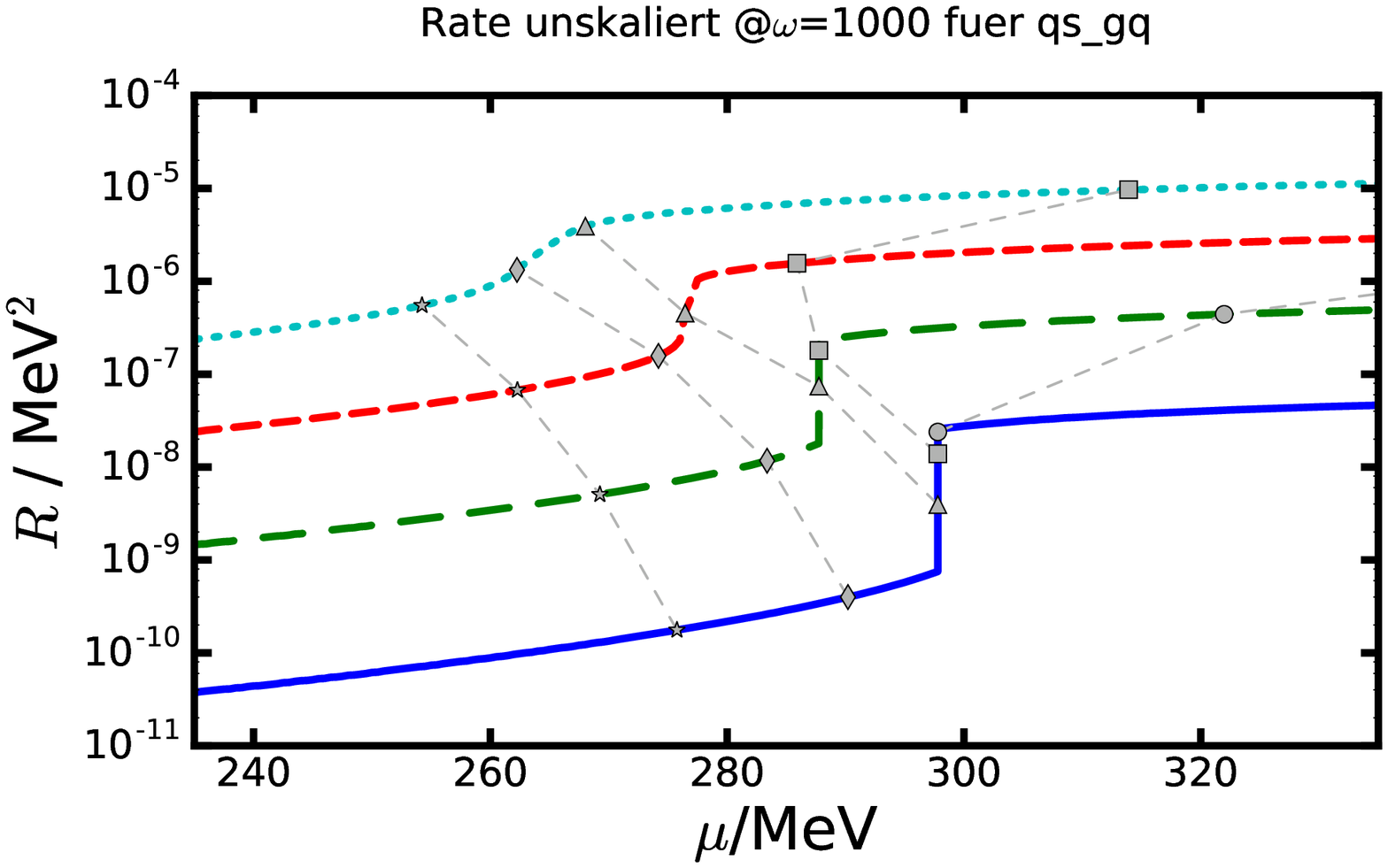}}
   \caption{Photon emission rates $R=\omega d^7 N / \dkdx$ at $\omega=1\GeV$ in the $\CEP$ region for the 
   Compton processes
   $q + \pi \rightarrow q + \gamma$ (upper row), 
   $q + \sigma \rightarrow q + \gamma$ (lower row).
   Left panels: contour plots of the rates in $\MeV^2$; right panels: rates at  constant temperature 
   $T/\MeV=$55, 65, 75 and 85 (bottom to top).
   The symbols denote the rates at isentropes with $s/n=$ 1.7 (dots), 2.1 (squares),
   2.5 (triangles), 2.9 (diamonds), 3.3 (stars) and the thin gray dashed curves are for guiding the eyes.
   Below, in Fig.~\ref{fig:isentropes}, these isentropes are displayed as black curves in the $T$-$\mu$ diagram.
   The solid white curves in the left plots depict the FOPT curves, and the white dashed line
   is an estimate of the crossover region based on the heat capacity. The dot depicts the position of the CEP, 
   numerically determined by the coordinates of the minimum of the $\sigma$ mass.}
   \label{fig:rate_qp_gq}
   \label{fig:rate_qs_gq}
   \label{fig:Compton_rate}
\end{figure}

\begin{figure}
   \centering
   {\includegraphics[trim=6mm 16mm 14mm 34mm,width=0.49\textwidth,clip]{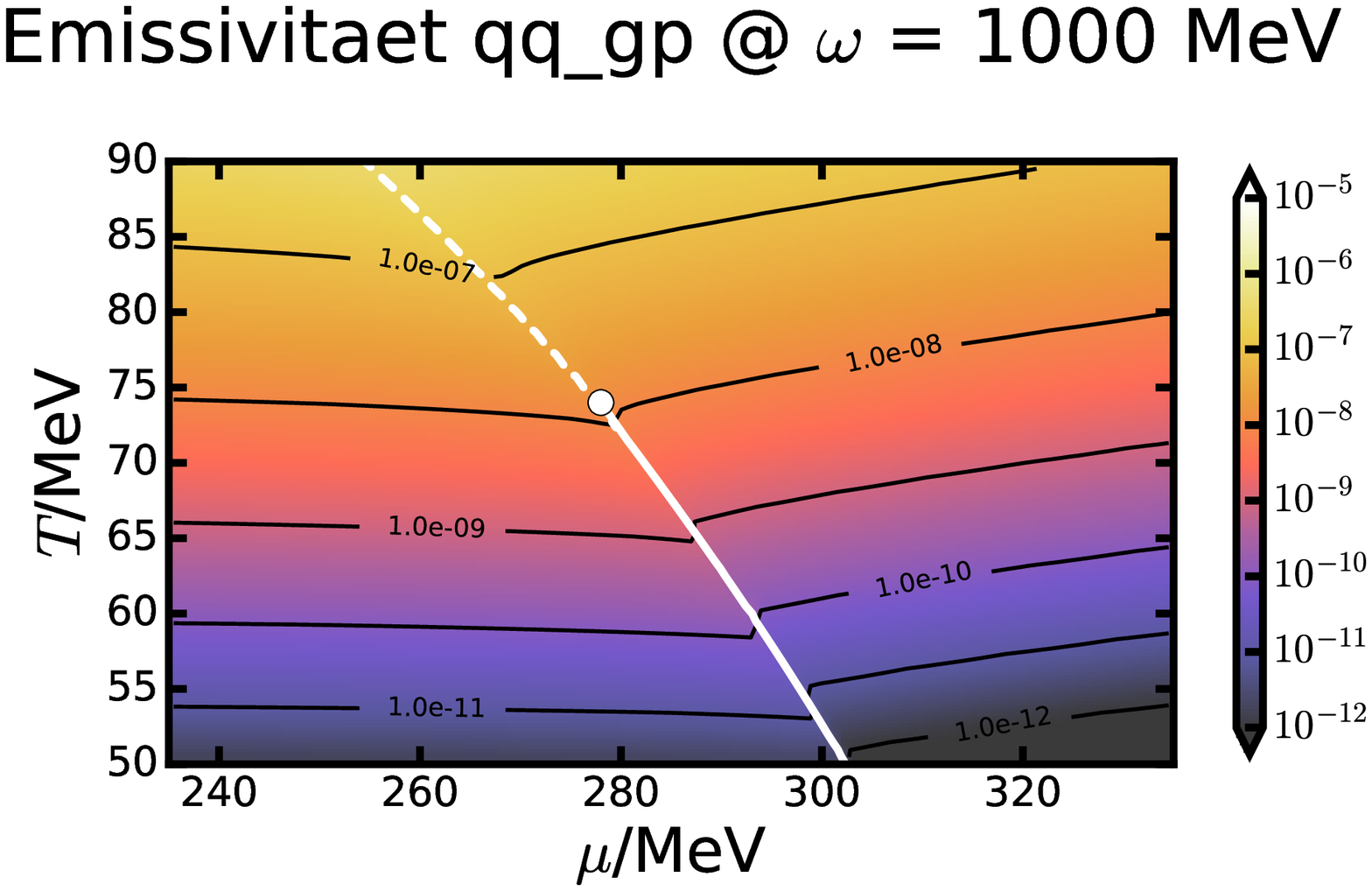}}
   {\includegraphics[trim=0mm 2mm 5mm 11mm,width=0.47\textwidth,clip]{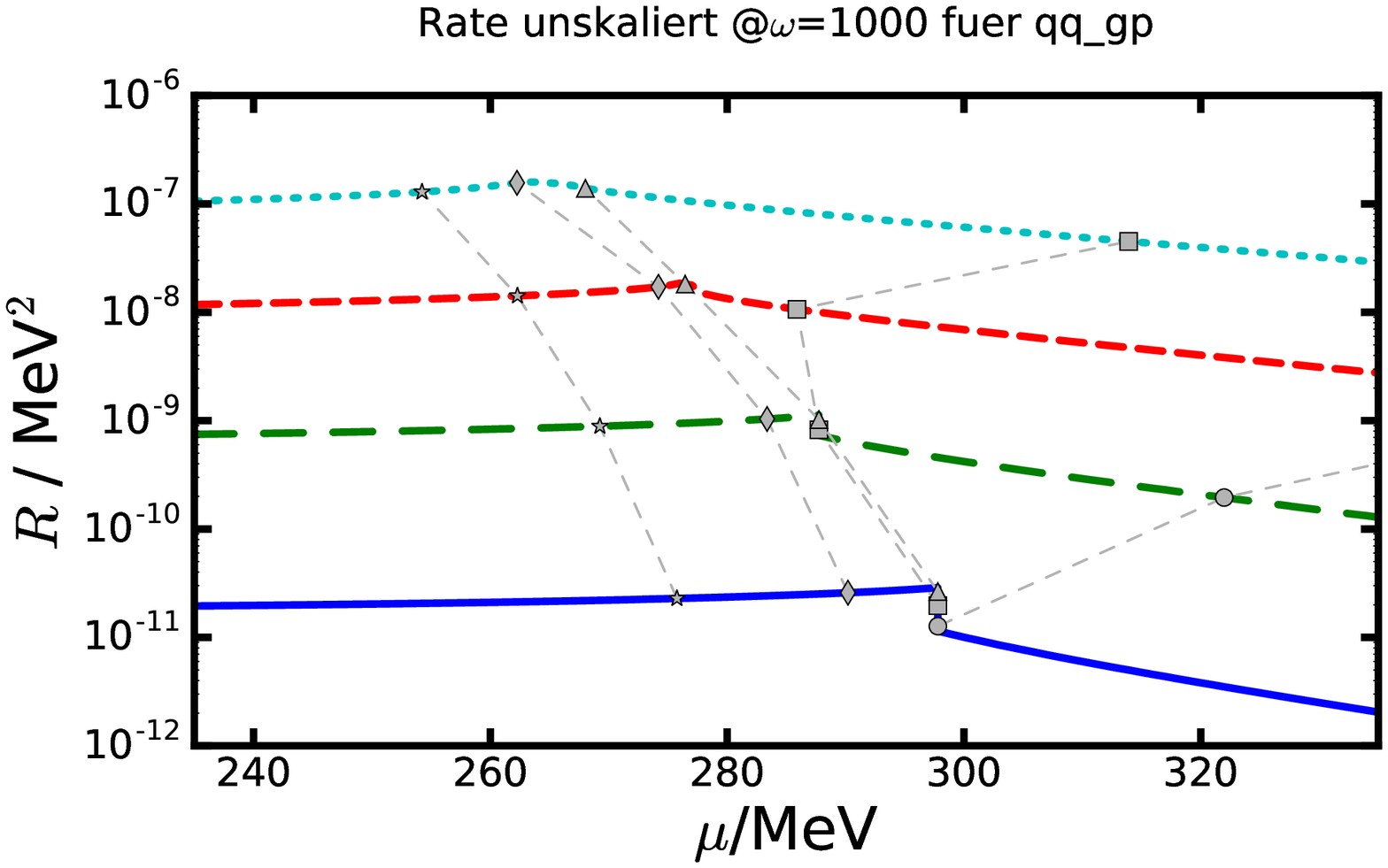}}\\
   {\includegraphics[trim=6mm 16mm 14mm 34mm,width=0.49\textwidth,clip]{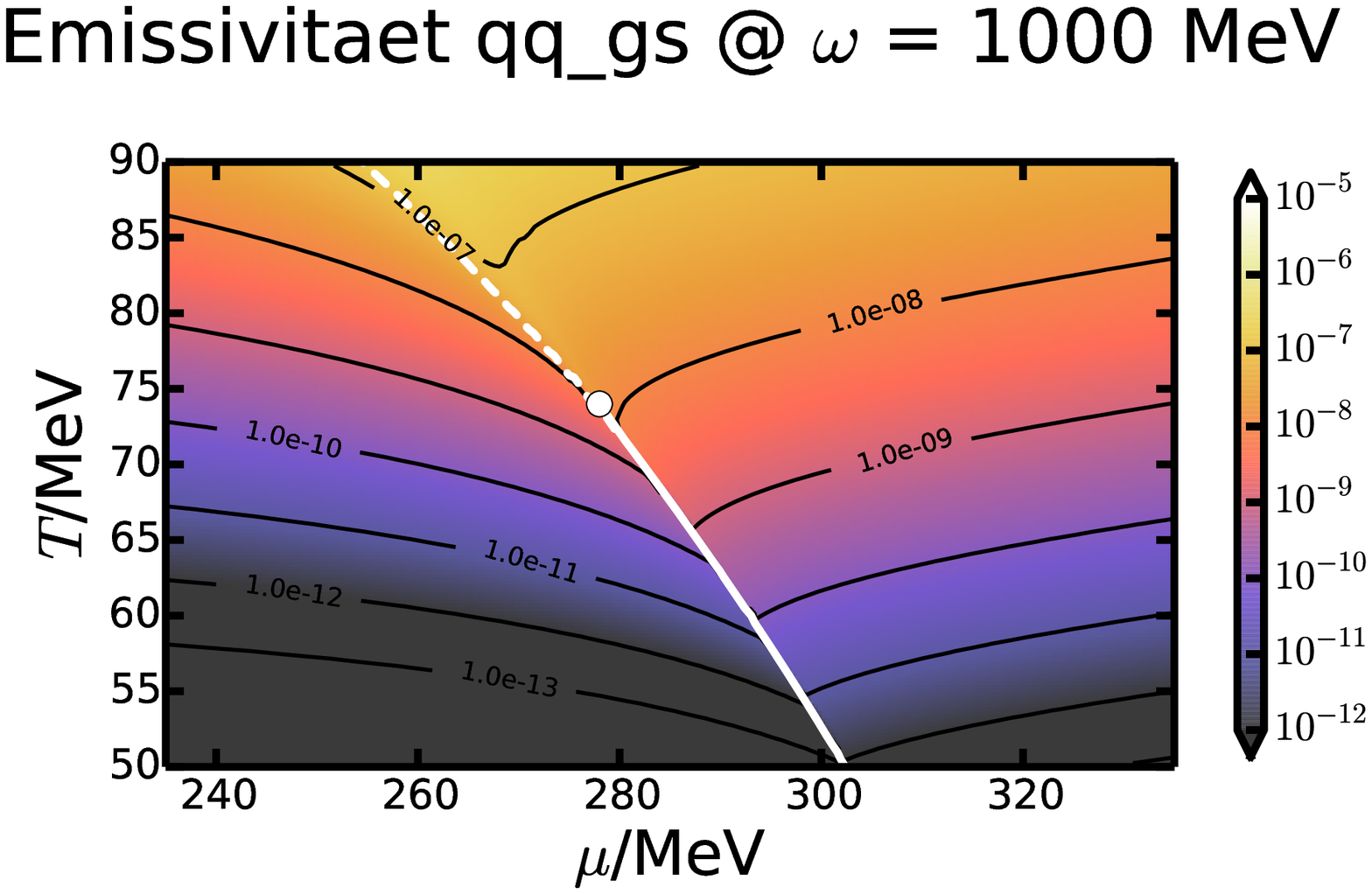}}
   {\includegraphics[trim=0mm 2mm 5mm 11mm,width=0.47\textwidth,clip]{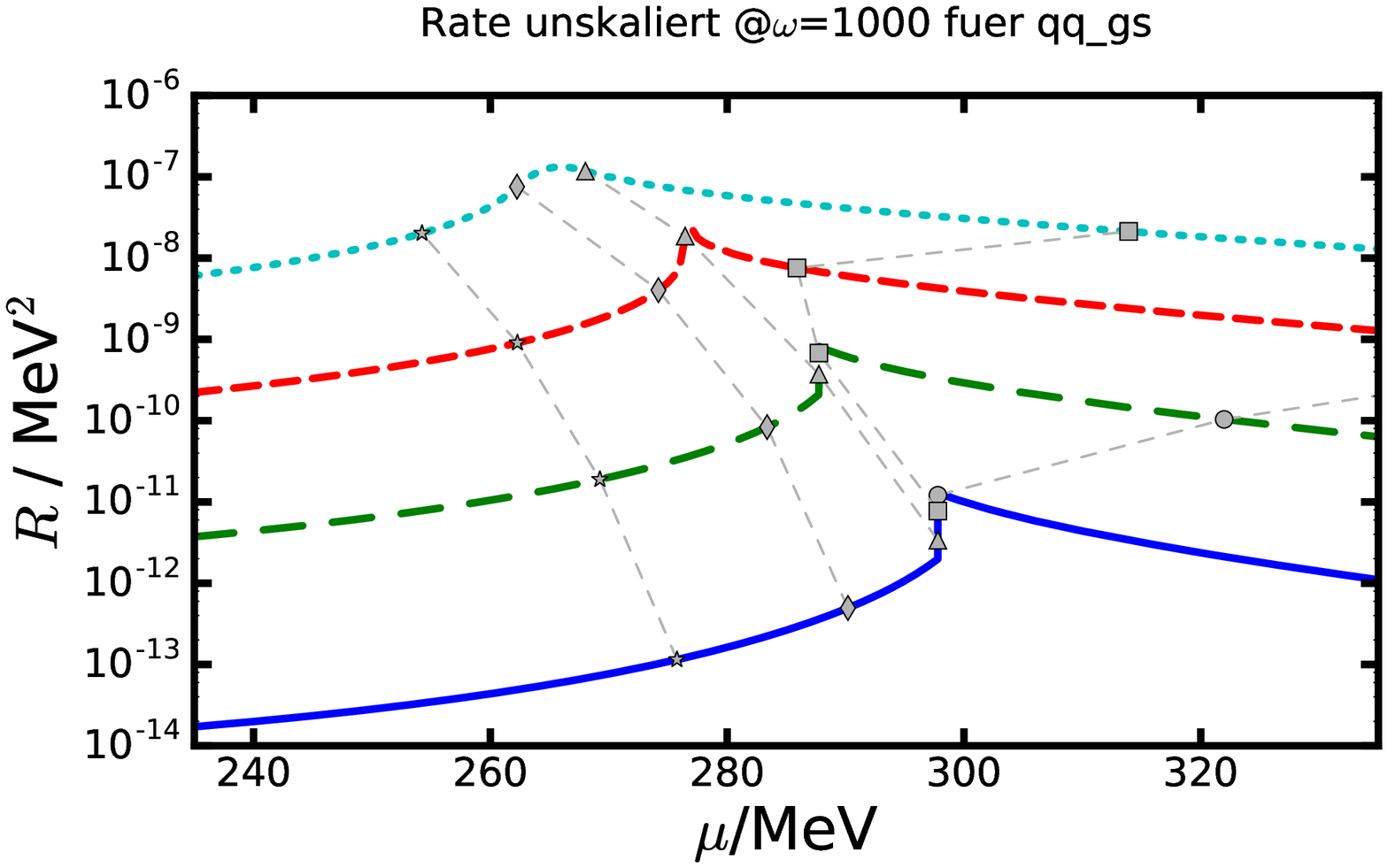}}
   \caption{As Fig.~\ref{fig:Compton_rate}, but for the annihilation processes
   $q + \bar q\rightarrow \pi + \gamma$  (upper row) and 
   $q + \bar q\rightarrow \sigma + \gamma$  (lower row).}
   \label{fig:rate_qq_gp}
   \label{fig:rate_qq_gs}
   \label{fig:Annihil_rate}
\end{figure}

\begin{figure}
   \centering
   {\includegraphics[trim=6mm 16mm 14mm 34mm,width=0.49\textwidth,clip]{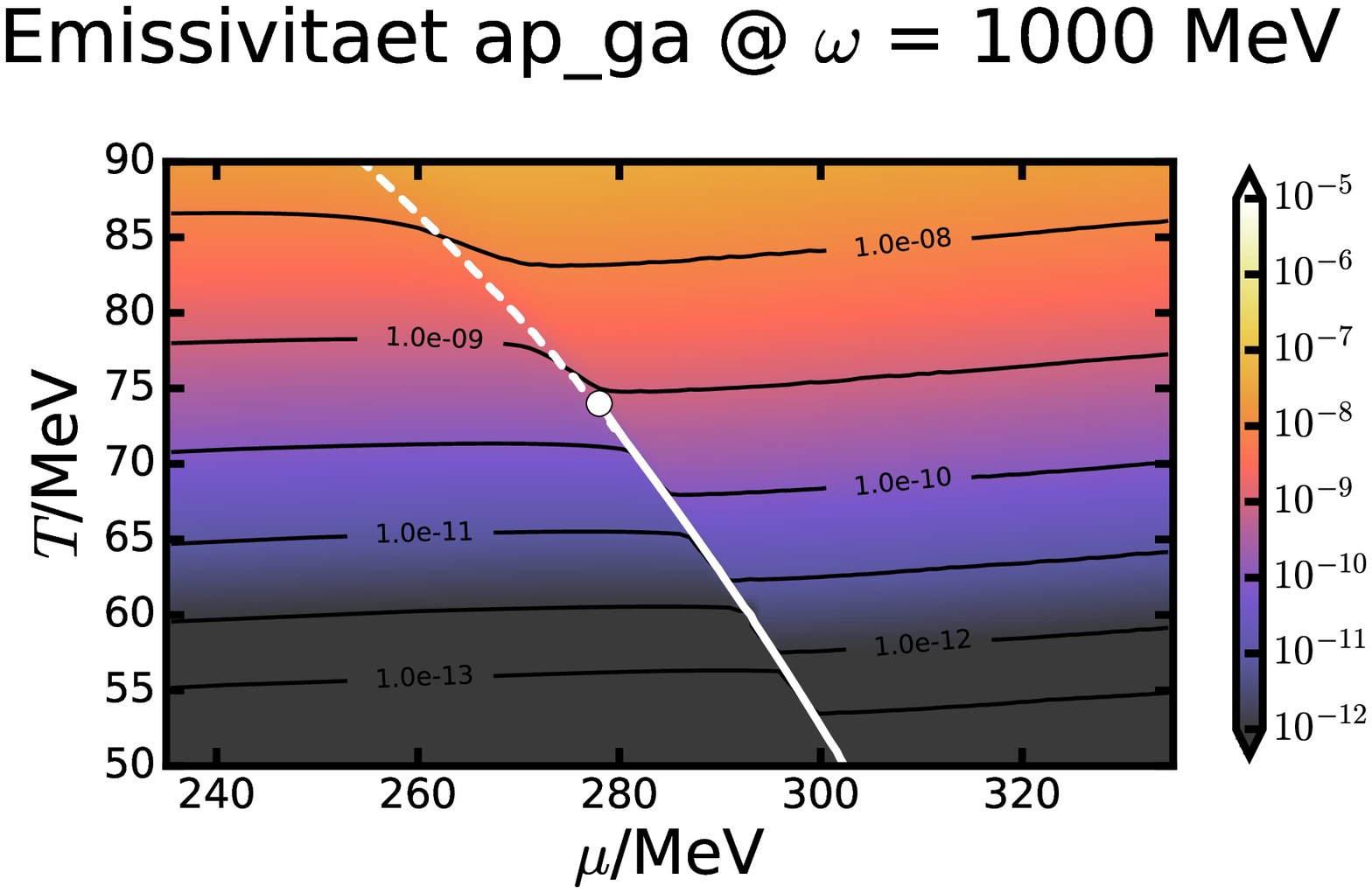}}
   {\includegraphics[trim=0mm 2mm 5mm 11mm,width=0.47\textwidth,clip]{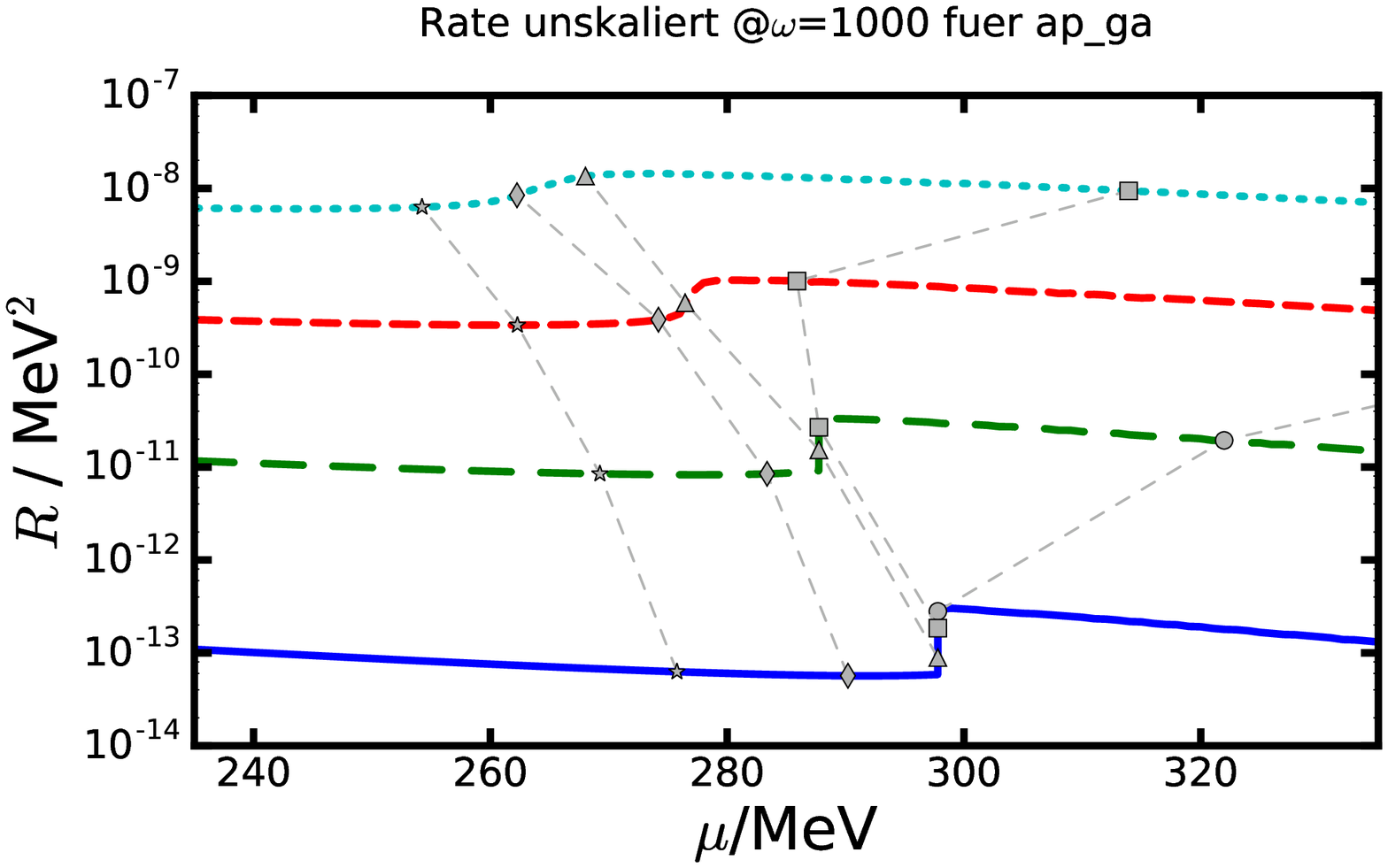}}\\
   {\includegraphics[trim=6mm 16mm 14mm 34mm,width=0.49\textwidth,clip]{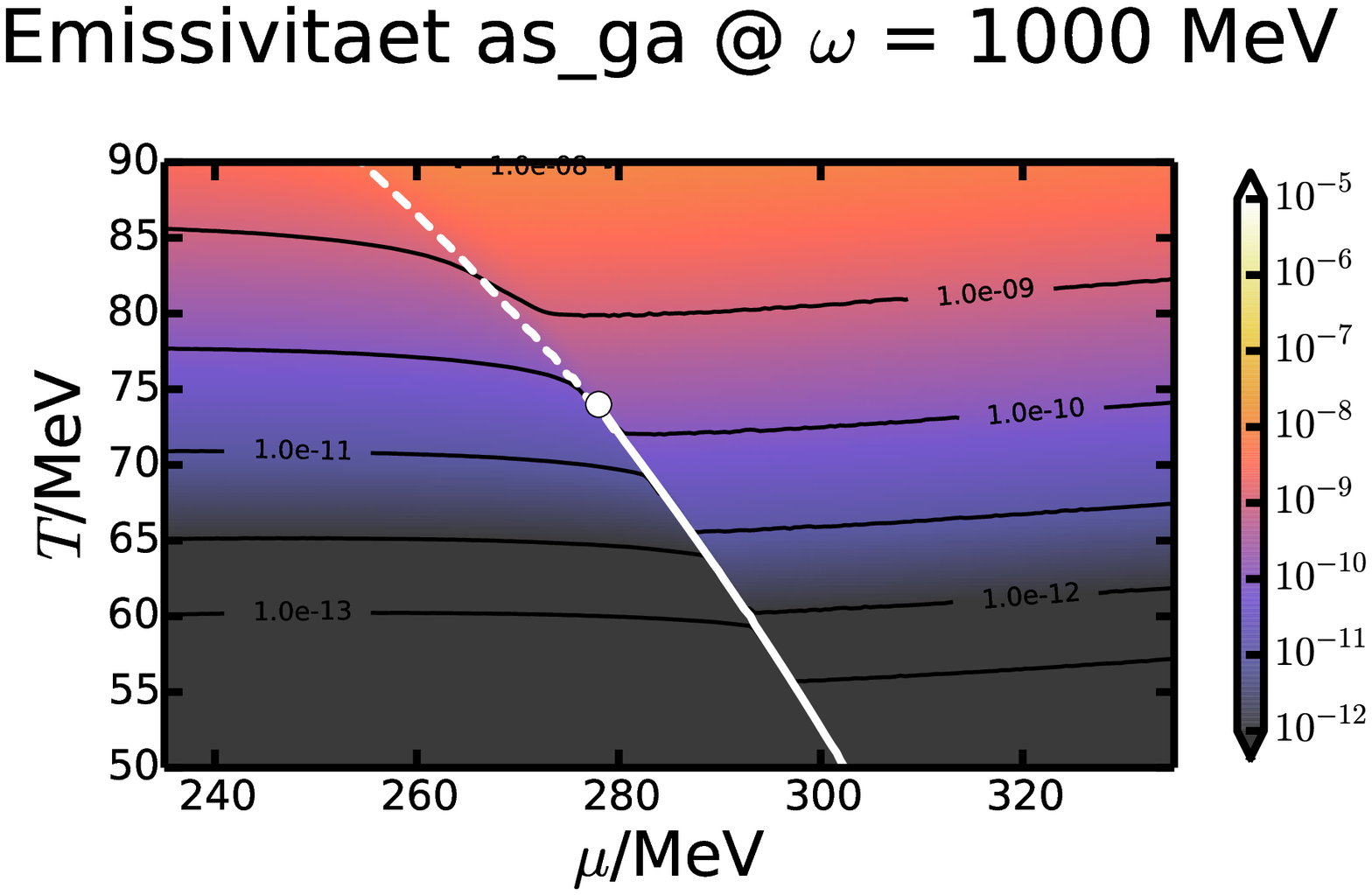}}
   {\includegraphics[trim=0mm 2mm 5mm 11mm,width=0.47\textwidth,clip]{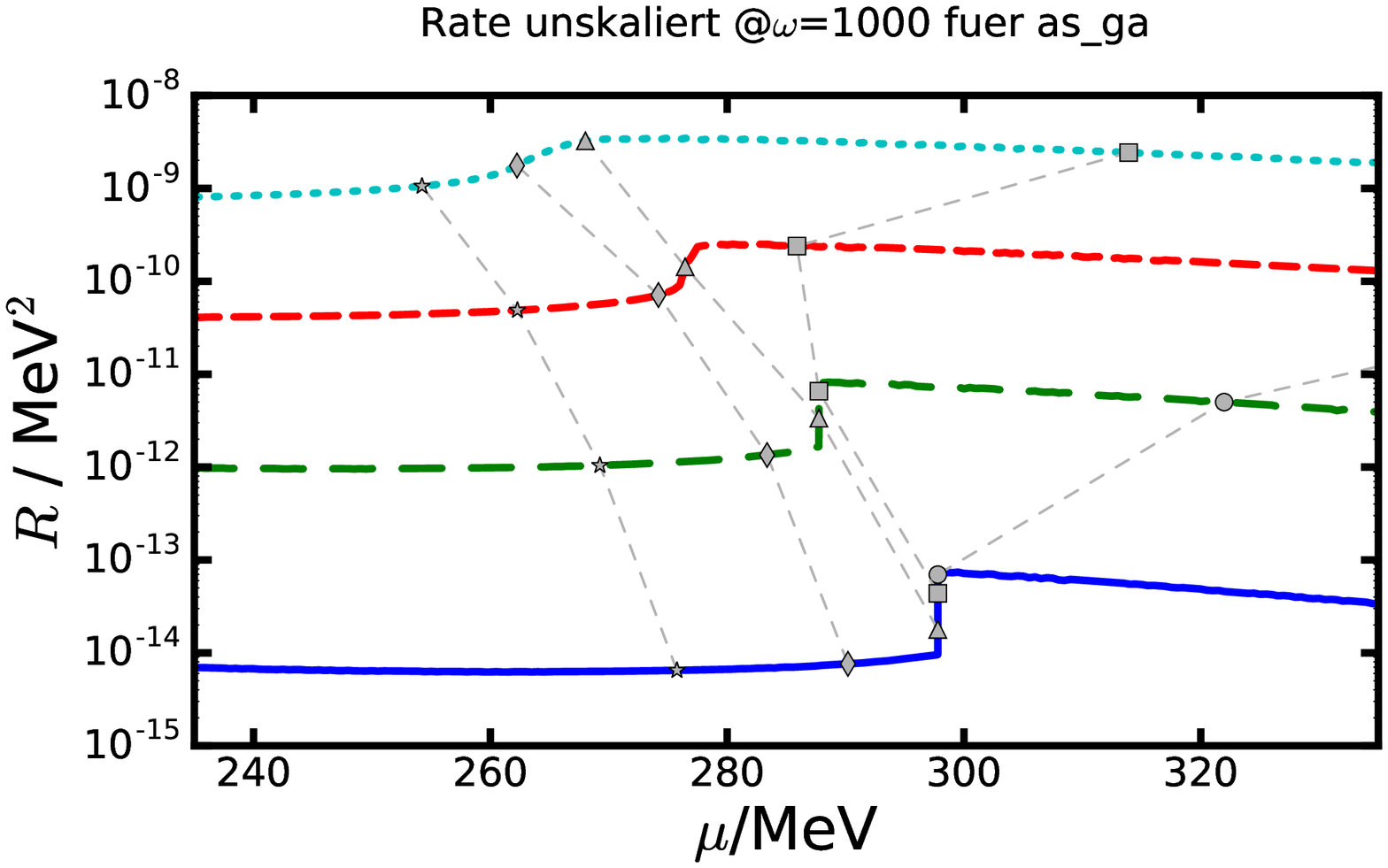}}
   \caption{As Fig.~\ref{fig:Compton_rate}, but for the anti-Compton processes
   $\bar q + \pi \rightarrow \bar q + \gamma$ (upper row) and
   $\bar q + \sigma \rightarrow \bar q + \gamma$ (lower row).}
   \label{fig:rate_ap_ga}
   \label{fig:rate_as_ga}
   \label{fig:AntiCom_rate}
\end{figure}

\begin{figure}
   \centering
   \includegraphics[trim=6mm 16mm 14mm 34mm,width=0.49\textwidth,clip]{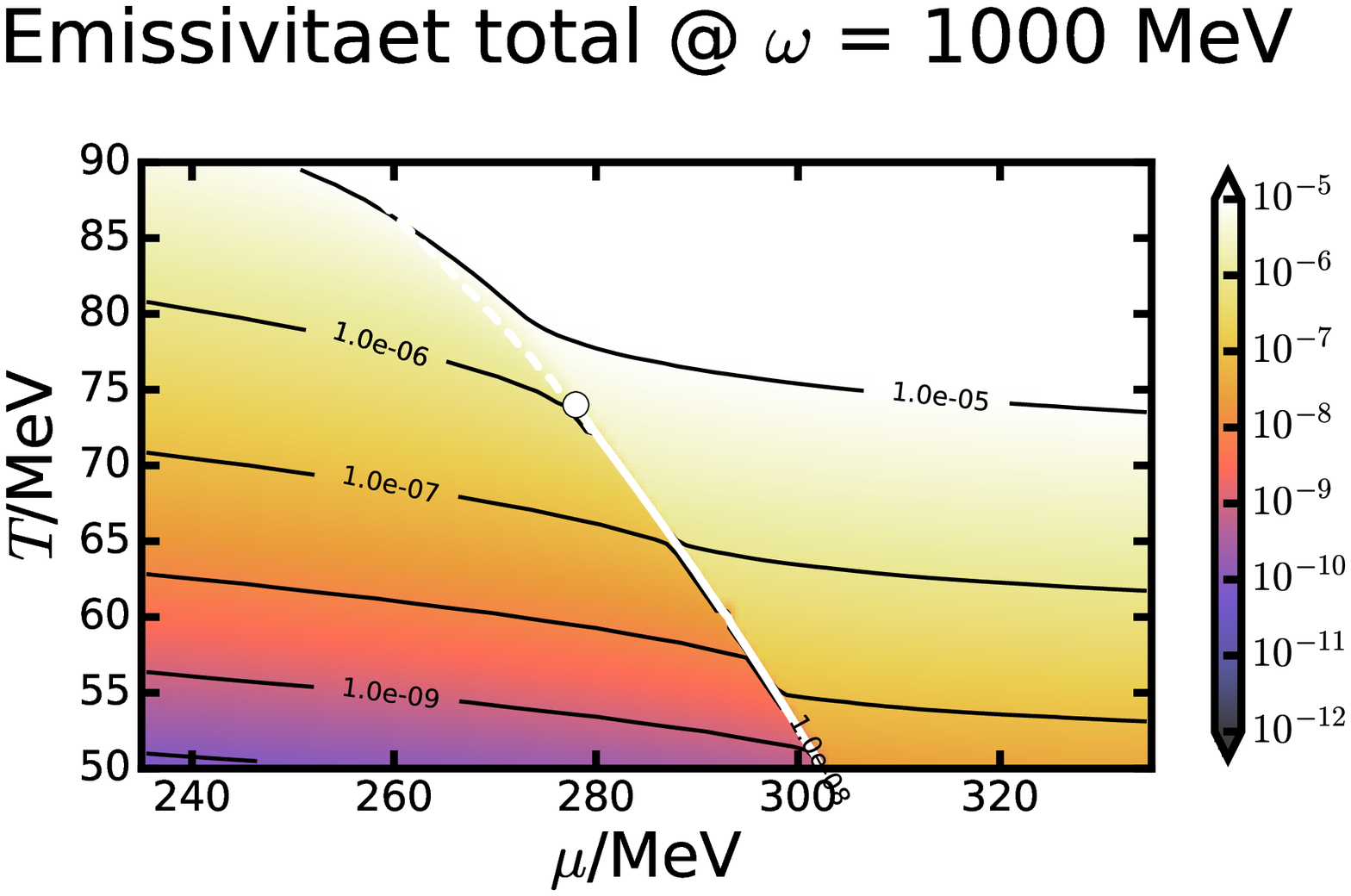}
   \includegraphics[trim=0mm 2mm 5mm 11mm,width=0.47\textwidth,clip]{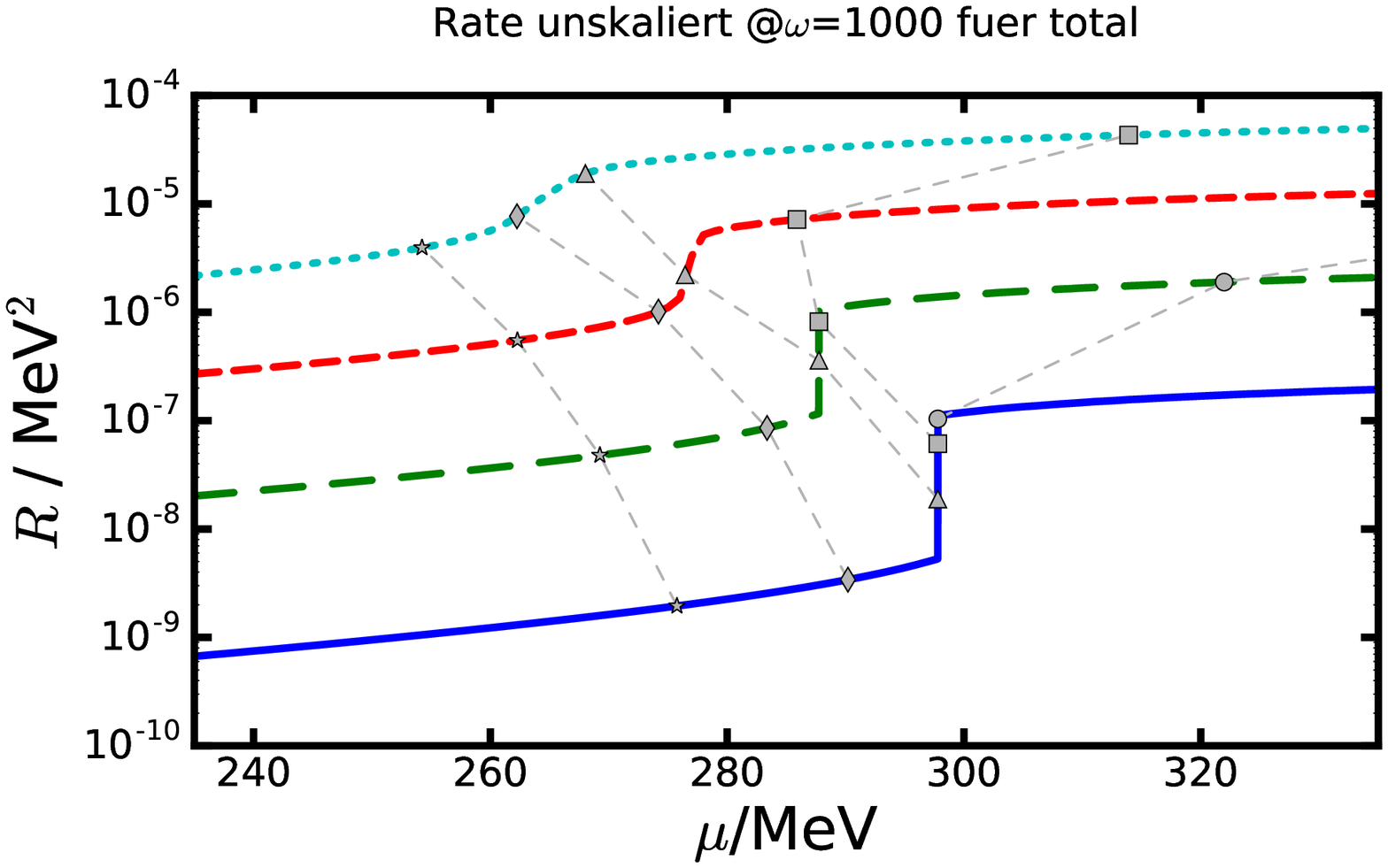}
   \caption{ 
           As Fig.~\ref{fig:Compton_rate}, but for the sum of the processes of \eqref{Compton_01}-\eqref{Annihil_01}.
           }
   \label{fig:total_rate}
\end{figure}

Our central question is to which extent the features of the phase diagram are reflected in the photon emission rates.
Therefore, we inspect the partial rates and their dependence on $T$ and $\mu$, \cf 
Figs.~\ref{fig:rate_qp_gq}-\ref{fig:rate_as_ga}. As mentioned in Section \ref{sec:diff_spectra} one sees a clear hierarchy
of the emission rates for the different types of processes in the chirally restored phase. In the chirally broken phase
the partial rate for the annihilation into $\pi$ and $\gamma$ is of similar size compared to the Compton processes,
although it is suppressed by a factor of $\exp\{-\mu/T\}$. The reason for that is the comparatively small mass of the 
pions as the pseudo-Goldstone modes, which compensates for this suppression. 
The jump of the rates at the FOPT increases with 
decreasing temperature and reaches a factor of about 50 at the lowest displayed temperatures in the plots 
(see Fig.~\ref{fig:Compton_rate}, right panels) for the
dominating processes, \ie the Compton processes, as well as for the total rate 
(see Fig.~\ref{fig:total_rate}, right panel).

Only the annihilation process with $\sigma$ mesons shows a non-monotonic behavior when scanning the partial rate along
curves with constant $T$ and varying $\mu$. This can be traced back to the effective $\sigma$ mass, which is relatively
small in a valley surrounding the phase contour (FOPT and crossover region) and minimal at the $\CEP$. Since the process
$q + \bar q \rightarrow \sigma + \gamma$
is the only one of the considered channels that is primarily be influenced by the $\sigma$ mass it is especially 
interesting for the search for $\CEP$ related features in the photon rates. However, the corresponding partial rate 
is strongly
reduced by the above mentioned suppression factor $\exp\{-\mu/T\}$ relative to the Compton channels which makes is practically
invisible in the total rate. This masking of the annihilation channels is expected to weaken drastically for parameter
sets showing a CEP closer to the $T$ axis.
\section{Discussion}
\label{sec:Discussion}
We have described an approximation scheme for calculating the thermodynamics and the effective masses of the fields
contained in the $\LSM$ Lagrangian beyond the standard mean field approximation. 
The presented
approximation scheme has been shown to be a consistent approximation for the determination of equilibrium thermodynamical 
properties and scattering or production rates. This makes the calculated meson masses applicable for $S$ matrix 
calculations of the production rate for photons for which we present the lowest order (in the electromagnetic as well
as the quark-meson coupling) results. 
Furthermore, we discuss in \ref{apdx_ThDyn} the influence of the model parameters on expansion 
properties provided by isentropes as well
as on landmarks (such as position of the $\CEP$, pseudocritical temperature at vanishing net density, general shape
of the transition (first order as well as crossover) curve) of the phase diagram finding that all of these can be 
understood and adjusted to desired values
with the help of two particular combinations of the parameters: the fermion vacuum mass $\mNuVac/3$ and the product of 
$\mSiVac$ and the vacuum expectation value of the sigma field $\vSiVac$ as well as their interplay, at least for explicit
symmetry breaking terms that are neither too small (\ie $\mPiVac\gtrsim 50\MeV$) to avoid being influenced by
the wrong chiral limit, nor too big (\ie $(\mPiVac/\mSiVac)^2 \ll 1$) in order for the $\LSM$ to invoke an 
only weakly broken chiral symmetry.

It turns out that in the intermediate photon energy range of $\omega \sim 1\GeV$ there are
still sizable effects in the photon production rates due to a FOPT.
Of course, for a firm result, many more emission channels have to be included 
(\eg in \cite{Liu:2007zzw}, the channel $\pi^+\pi^-\rightarrow \sigma/\rho \rightarrow \pi^+\pi^-\gamma$ is identified to
be important in the soft photon regime and in \cite{Linnyk:2015tha} $2\rightarrow3$ processes, such as 
meson-meson and meson-baryon bremsstrahlung,
are found to be of great importance) and the effect of inclusion of higher order
terms in the quark-meson coupling has to be checked as well as the effect of including further fluctuations 
(as done, \eg in \cite{Tripolt:2013jra} within the FRG framework). 
In a previous work \cite{Wunderlich:2015rwa} we showed that the dominant effect on the photon rates stems from
the mass variations and the explicit $\mu$ dependencies of the distribution function; in other words it is of kinematical
origin. This leads us to the conjecture that the position and size of the discontinuities in the photon rate
is a robust feature and could probably provide a tool suitable for the detection of a chiral FOPT
in HIC experiments.
\section{Summary}
\label{sec:Summary}
In summary, we employ here a quark-meson model with linearized fluctuations of the meson fields, which displays
the onset of a curve of FOPTs at a (albeit imperfect) $\CEP$. The thermodynamics
has been elaborated in previous works 
\cite{Mocsy:2004ab,Bowman:2008kc,Ferroni:2010ct,Wunderlich:2015rwa,Wunderlich:2014cia}. We couple the 
pertinent degrees of freedom to the electromagnetic field to evaluate the photon emission rates over the phase diagram,
in particular the impact of the FOPT. The chain of approximations is pointed out to arrive at 
emission rates in the form of kinetic theory expressions being consistent with the thermodynamics.
To this end it is necessary to go beyond the mean field approximation, because in such an approach the mesons 
are no dynamic fields which is conceptually inconsistent with their usage in $S$ matrix calculations. 
The first step in a path integral approach beyond mean fields is the inclusion of the lowest order fluctuations,
which we achieve by the quadratic approximation of the effective mesonic potential. Our calculation differ from
that in \cite{Mocsy:2004ab,Bowman:2008kc,Ferroni:2010ct} by the inclusion of photons and the source terms for
all fields (see \ref{apdx_ThDyn}). The source terms make it possible to derive thermodynamics and $S$ matrix
elements on the same footing thus achieving consistency between both. Especially we can pin down the correct quark mass
parameter for the calculation in the kinetic theory framework, which was not possible in previous works.

Due to the tight coupling of emissivities in lowest-order tree-level diagrams and thermodynamics,
it happens that individual channels of photon producing processes map out the phase diagram. The emission
rates are determined essentially by the effective masses of the involved field modes.
While soft photons are either suppressed by finite temperature effects or enhanced by infrared divergencies 
of the matrix elements, the hard photons display the usually expected exponential shapes.
Chiral restoration as degeneracy of pion and sigma effective masses causes also a degeneracy of the partial rates in
the restored phase.
The hard photon rates obey in the chirally restored phase for $\mu/T\gtrsim 1$ the following hierarchy: 
The rates from Compton-processes are larger
than those from annihilations, which in turn are larger than those from anti-Compton processes.

We supplement our study by a discussion of the parameter dependence of the $\CEP$ coordinates and the
location of the FOPT curve as well as the pattern of isentropic curves relevant for adiabatic
expansion paths in the phase diagram (see \ref{apdx_ThDyn}).

Finally, we mention that our investigation should be considered as a case study, not mimicking QCD features sufficiently
adequate. Beyond the impact of vacuum fluctuations, 
the involved degrees of freedom mistreat (i) at low temperatures the nucleons and their incompressibility,
as well as the other known hadronic states needed to saturate the equation of state known from QCD, 
and
(ii) at high temperatures the explicit gluon degrees of freedom.
Nevertheless, we stress again that a seemingly universal emissivity must not be combined with an ad hoc
assumed thermodynamics/phase structure, however both issues must be dealt with in a consistent manner.
\section*{Acknowledgments}
   We thank J. Randrup, V. Koch, F. Karsch, K. Redlich, M.I. Gorenstein, S. Schramm, H. St\"ocker, B.J. Schaefer, 
   B. Friman, R. Rapp, H. van Hees and C. Gale 
   for enlightening discussions of phase transitions in nuclear matter and photon emission. 
   The work is supported by BMBF grant 05P12CRGH and TU Dresden graduate academy scholarship grant 
   F-003661-553-62A-2330000.
\appendix
\section{A few formal details}
\subsection{Derivative expansion}
\label{apdx_deriv_exp}
In the case of a $\phi^4$ theory the method is explained in \cite{Fraser:1984zb,Aitchison:1985pp}. For convenience we
outline it here and apply it to the theory at hand.
The quantity we want to approximate is
\begin{align}
   \Omega_\qq &= -\Tr\ln \left[ \big(G^0_\psi(\sigma, \vec\pi)\big)^{-1} \right]
\end{align}
with $\big(G^0_\psi(\sigma, \vec\pi)\big)^{-1}$ defined according to \eqref{Lagrangian_02}.
Formally we can expand
\begin{align}
   \Omega_q &= -\Tr\ln \left[ i\slashed \partial - g(\sigma + i\gamma^5\vec\tau\vec\pi) \right]\\
              &= -\Tr\ln \left[ i\slashed \partial
                                \left(1-\frac{1}{i\slashed\partial}g(\sigma + i\gamma^5\vec\tau\vec\pi)\right)\right]\\
              &= -\Tr\ln \slashed p - \Tr\ln(1 + \slashed p^{-1} M)\\
              &\approx -\Tr\ln \slashed p - \Tr\left[ \slashed p^{-1} M \right] + \frac12\Tr\left[ \slashed p^{-1} M \slashed p^{-1}M\right] - \dots,\label{Log_exp_02}
\end{align}
were we used the shortcut $M = g(\sigma + i\gamma^5\vec\tau\vec\pi)$.
Applying $(\slashed p)^{-1} =\slashed p/p^2$ and the fact that the trace of an odd number of Dirac matrices vanishes
we see that only powers of $ \slashed p^{-1} M \slashed p^{-1}M$ remain in the sum (besides the $\ln \slashed p$-term).
Using
\begin{align}
   \slashed p^{-1} M \slashed p^{-1} M 
      =& \frac{\slashed p}{p^2}g(\sigma + i\gamma^5\vec\tau\vec\pi)\frac{\slashed p}{p^2}g(\sigma + i\gamma^5\vec\tau\vec\pi)\\
      =& \frac{\slashed p}{p^2}\gamma^\mu g(\sigma - i\gamma^5\vec\tau\vec\pi)\frac{p_\mu}{p^2} g(\sigma + i\gamma^5\vec\tau\vec\pi)
\end{align}
and
\begin{align}
   \phi(x)p_\mu = p_\mu \phi(x) + [\phi(x),p_\mu] = p_\mu \phi(x) - i\partial_\mu \phi(x),
\end{align}
for any field $\phi(x)$ we arrive at
\begin{align}
  \slashed p^{-1} M \slashed p^{-1} M 
     =&   \frac{\slashed p}{p^2}\slashed p g(\sigma - i\gamma^5\vec\tau\vec\pi)
          \frac{1}{p^2}g(\sigma + i\gamma^5\vec\tau\vec\pi)
       -i\frac{\slashed p}{p^2}\Big(\slashed\partial g(\sigma - i\gamma^5\vec\tau\vec\pi)\Big)
          \frac{1}{p^2} g(\sigma + i\gamma^5\vec\tau\vec\pi).\label{pMpM_01}
\end{align}
Employing the operator identity (for $A$ invertible)
\begin{align}
   [A^{-1}, B] &= -A^{-2}[A,B] - A^{-3}[A,[A,B]] - A^{-4}[A,[A,[A,B]]] - \dots\label{comm_ident_01}
\end{align}
with $A= p^2$ and $B=\sigma,\pi$ the $1/p^2$ term in \eqref{pMpM_01} can be commuted to the left.
The nested commutators in \eqref{comm_ident_01} are computed by utilizing recursively the identity
\begin{align}
   [p^2,\phi] =& \square \phi + 2i p^\mu\partial_\mu \phi.\label{comm_ident_02}
\end{align}
Inspecting $\eqref{comm_ident_02}$ one sees that each commutator with $p^2$ contributes at least
one derivative of $\phi$ leading to the observation that terms in \eqref{comm_ident_01} with $n$ commutators imply
at least $n$ derivatives of the meson fields. Thus, we find
\begin{align}
   [p^{-2}, \sigma \text{ or }\pi] = 0 + \Ord{\partial \sigma,\partial \vec \pi}
\end{align}
leading to
\begin{align}
   \slashed p^{-1} M \slashed p^{-1} M 
    =& \frac{1}{p^2}g(\sigma - i\gamma^5\vec\tau\vec\pi)
          g(\sigma + i\gamma^5\vec\tau\vec\pi) + \Ord{\partial \sigma,\partial \vec \pi}\\
    =& \frac{1}{p^2}m_q^2 + \Ord{\partial \sigma,\partial \vec \pi}
\end{align}
with $m_q^2 = g^2(\sigma^2+\vec\pi^2)$.
Taking only zero derivative terms, the higher powers of $\slashed p^{-1} M \slashed p^{-1} M$ in the expansion 
\eqref{Log_exp_02} result in 
\begin{align}
   (\slashed p^{-1} M \slashed p^{-1} M)^n = \left(\frac{1}{p^2}\right)^n \left(m_q^2\right)^n\mathbb{1}_D +  \Ord{\partial \sigma,\partial \vec \pi}
\end{align}
with $\mathbb{1}_D$ denoting the unity matrix in Dirac space.
Then the complete expansion \eqref{Log_exp_02} gives
\begin{align}
   \Omega_q =&\approx -\Tr\ln \slashed p 
                       - \frac12\Tr\left[ \frac{m_q^2}{p^2} \mathbb{1}_D\right] 
                       - \frac14\Tr\left[ \left(\frac{m_q^2}{p^2}\right)^2 \mathbb{1}_D\right]-\dots 
                       + \Ord{\partial\sigma,\partial\pi^a}.
\end{align}
It can easily be checked that this is exactly the expansion of a noninteracting Fermi gas with mass $m_q$ 
\begin{align}
   -\Tr\ln\Big[\slashed p -m_q\Big] = -\Tr\ln\slashed p 
                               - \sum_n \frac{1}{2n}\Tr\Bigg[\left(\frac{m_q^2}{p^2}\right)^n\mathbb{1}\Bigg],
\end{align}
thus verifying \eqref{det_G_q_01}.
\subsection{Inverting perturbed matrices}
\label{apdx_invert_mtrx}
We apply
\begin{align}
   M^{-1} =& M_0^{-1} \sum_{n=0}^\infty(-\Delta_M M_0^{-1})^n, & M =& M_0 + \Delta_M \label{matrix_inversion_01}
\end{align}
valid for invertible matrices $M$ and $M_0$. 
A heuristic derivation of \eqref{matrix_inversion_01} can be obtained by noting
\begin{align}
   M^{-1} = (M_0 + \Delta_M)^{-1} = M_0^{-1}(1 - (-M_0\Delta_M))^{-1} 
\end{align}
which is then written as a geometric series, leading to \eqref{matrix_inversion_01}.
With $M^a_b \equiv M(x_a,x_b)$ this relations can 
be reformulated for the continuum limit in the language of functional derivatives with the only 
changes being $\partial/\partial \phi_i\rightarrow \delta/\delta\phi(x)$
and the matrix multiplication replaced by an integral $A^a_b B^b_c\rightarrow \int db A(x_a,b)B(b,x_c)$. 
\subsubsection{Application to the photon propagator}
Setting
$M(z,z') \equiv \big(\Gph[(z,z')]{\mu\nu}\big)^{-1} 
 = \bar G^\gamma_{\mu\nu}(z,z')^{-1} + \big[- e^2 \pi^+(z)\pi^-(z)g_{\mu\nu}\big]\delta(z-z')$ 
one gets 
\begin{align}
  \Gph[(z,z')]{\mu\nu} =& \bar G^\gamma_{\mu\nu}(z,z') + \int d^4x\bar G^\gamma_{\mu\rho}(z,x)\big[e^2 \pi^+(x)\pi^-(x)g^{\rho\kappa}\big]\bar G^\gamma_{\kappa\nu}(x,z') 
                                   + \Ord{e^4},\label{PhotonProp_02}
\end{align}
which is applied in \eqref{em_contrib_02}.
\subsubsection{Application to the quark propagator}
Setting
$M(z,z') \equiv \big(\Gqu[(z,z')]\big)^{-1} = \Gquvev[(z,z')]^{-1} + \big[-g \Delta(z) -gi\gamma^5\tau_a\pi^a(z)\big]\delta(z-z')$ 
one obtains 
\begin{align}
   \Gqu[(z,z')] =& \Gquvev[(z,z')] - \int d^4x\Gquvev[(z,x)]A(x)\Gquvev[(x,z')] \notag\\
                 &                 - \iint d^4xd^4y\Gquvev[(z,x)]
                                                   A(x)\Gquvev[(x,y)]
                                                   A(y)\Gquvev[(y,z')]\label{quark_prop_02}\\
                 &                 - \iint d^4xd^4y\Gquvev[(z,y)]
                                                   A(y)\Gquvev[(y,x)]
                                                   A(x)\Gquvev[(x,z')]
                                   + \Ord{\Delta^3,\vec \pi^3},\notag\\
   A(z)        =& -g \Delta(z) -gi\gamma^5\tau_a\pi^a(z).
\end{align}
Replacing on both sides of the equation $\Delta(z)$ by $\delta / \delta\eta_\sigma(z)$ and 
$\pi^a(z)$ by $\delta / \delta\eta_{\pi^{-a}}(z)$ one arrives at
\begin{align}
   \dGqu[(z,z')] =& \Gquvev[(z,z')] - \int d^4x\Gquvev[(z,x)]\hat A(x)\Gquvev[(x,z')] \notag\\
                 &                 - \iint d^4xd^4y\Gquvev[(z,x)]
                                                   \hat A(x)\Gquvev[(x,y)]
                                                   \hat A(y)\Gquvev[(y,z')]\label{quark_prop_04}\\
                 &                 - \iint d^4xd^4y\Gquvev[(z,y)]
                                                   \hat A(y)\Gquvev[(y,x)]
                                                   \hat A(x)\Gquvev[(x,z')]
                                   + \Ord{\frac{\delta^3}{\delta\eta_\sigma^3},\frac{\delta^3}{\delta\eta_\pi^3}},\notag\\
   \hat A(z)        =& -g \frac{\delta}{\delta\eta_\sigma(z)} -gi\gamma^5\tau_a\frac{\delta}{\delta\eta_{\pi^{-a}}(z)},
\end{align}
which is used in Section \ref{sec:photon_prop}.
\section{Thermodynamics and phase structure}
\label{apdx_ThDyn}
\subsection{Thermodynamics}
Setting all sources to zero in \eqref{GenFunc_01} transforms $S_\eta$ formally into the grand canonical partition 
function $Z$. As our goal is to study systems much smaller than the mean free path of photons (which is a reasonable
assumption in the context of HICs) the photons do not contribute to the pressure. Thus we remove 
all terms containing the photon field $A$ from \eqref{GenFunc_01}, which corresponds to setting zero the electromagnetic
coupling $e$ (explicitly and implicitly in $J^\mu_\gamma$) as well as removing $\det G_\gamma$ from \eqref{GenFunc_07}.
Then we get
\begin{align}
      Z =&\sqrt{\det \Gpi}^3\sqrt{\det \Gsi}\exp\left\{-\int d^4x\langle U_\eff\rangle
           +\frac12 m_\pi^2\langle\vec\pi^2\rangle +\frac12m_\sigma\langle\Delta^2\rangle\right\}\label{part_funct_01}.
\end{align}
As $\langle U_\eff\rangle$,$\langle\vec\pi^2\rangle$,$\langle\Delta^2\rangle$ and $m_{\sigma,\pi}$ do not 
depend on the space-time coordinates, the integration in the exponent yields a factor of the Euclidean volume $V\beta$.
For the grand canonical potential $\Omega(T,\mu) = -p(T,\mu) = (\beta V)^{-1}\ln Z$ one gets
\begin{align}
      \Omega =&\frac{3}{2}\ln{\det \Gpi}+\frac{1}{2}\ln{\det \Gsi}-\langle U_\eff\rangle
           -\frac12 m_\pi^2\langle\vec\pi^2\rangle -\frac12m_\sigma\langle\Delta^2\rangle\label{ThD_pot_01}.
\end{align}
Applying $\ln\det G_{\pi,\sigma}$ = $\Tr\ln G_{\pi,\sigma}$ and using standard techniques \cite{Kapusta:2006pm} 
for solving these functional traces one arrives at
\begin{align}
      \Omega =&\Omega_\pi+\Omega_\sigma+\langle U\rangle + \langle\Omega_\qq\rangle
           -\frac12 m_\pi^2\langle\vec\pi^2\rangle -\frac12m_\sigma\langle\Delta^2\rangle\label{ThD_pot_02},\\
      \Omega_\pi    =& \frac{3}{3(2\pi)^3}\int dp^3 \frac{p^2}{E_\pi}(1 + n_B(E_\pi),)\\
      \Omega_\sigma =& \frac{1}{3(2\pi)^3}\int dp^3 \frac{p^2}{E_\sigma}(1 + n_B(E_\sigma)),\\
      E_{\pi,\sigma}^2 =& m_{\pi,\sigma}^2 + \vec p^2
\end{align}
and $\Omega_\qq$ according to \eqref{fermion_pot_01} in agreement with \cite{Mocsy:2004ab,Bowman:2008kc,Ferroni:2010ct}.
From the thermodynamic potential the thermodynamic quantities 
(energy density, net quark density, entropy density, susceptibilities, etc.)
follow by differentiation. The explicit formulas have been worked out in \cite{Mocsy:2004ab,Bowman:2008kc,Ferroni:2010ct}.
\subsection{Impact of model parameters on the phase diagram}
\label{sec:PhaseDiag}
\begin{figure}
   \centering
   {\includegraphics[clip=true,trim=10mm 2mm 9mm 20mm,width=0.68\textwidth]{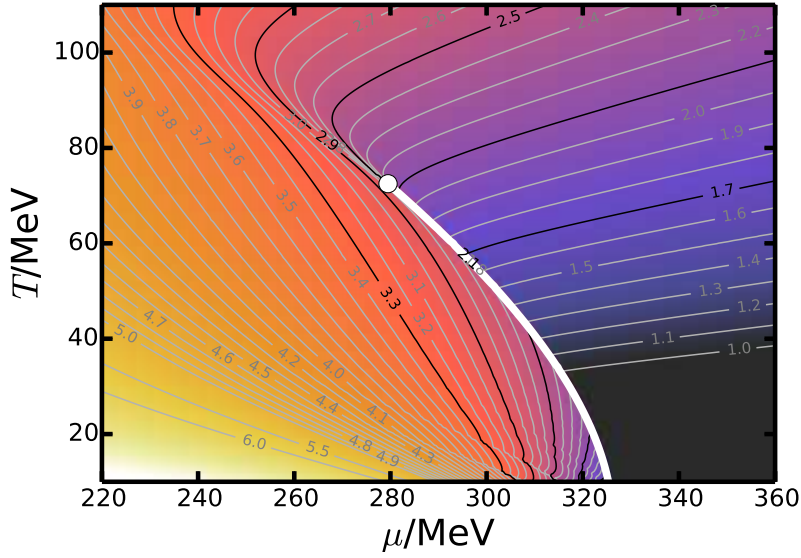}}
   \caption{Contour plot of entropy per quark over the phase diagram for parameter set A in Tab.~\ref{tab:parameters}. 
            Also plotted are isentropes (gray and black curves) with their
            respective $s/n$ values indicated. The black isentropes are those marked in the right panels of 
            Figs.~\ref{fig:Compton_rate}-\ref{fig:total_rate}.
            The solid white curves depicts the coexistence curve for the FOPT, and the white blob depicts the $\CEP$.
            }
   \label{fig:isentropes}
\end{figure}

For the sake of an easy comparison with literature (\cf \cite{Carignano:2016jnw} for parameter fixings when including 
vacuum fluctuations) we choose the parameters as in
\cite{Mocsy:2004ab,Bowman:2008kc,Ferroni:2010ct,Wunderlich:2015rwa}, corresponding to parameter set A in
Tab.~\ref{tab:parameters}.
(The effect of other parameter choices is discussed in 
\cite{Wunderlich:2015rwa} for
different $\sigma$ vacuum mass fixings, in \cite{Bowman:2008kc} for different $\pi$ vacuum masses and
in \cite{Schaefer:2008hk} for the three flavor model w.r.t. explicit symmetry breaking parameters and the $\sigma$ mass.)
The structure of the phase diagram is conform with expectations spelled out in \cite{Wunderlich:2016aed}:
Isentropic curves as indicators of the paths of fluid elements during adiabatic expansion "go through" the phase 
border curve. The type IA FOPT (in the nomenclature of \cite{Wunderlich:2016aed}) is realized by our model with 
parameter set A.
Such a choice leads to the phase diagram depicted in Fig.~\ref{fig:isentropes} with the CEP coordinates being 
$T_\CEP = 74\MeV$
and $\mu_\CEP = 278\MeV$.
Typically (and in fact in all of the above cited references) one or more parameters of the Lagrangian 
or the vacuum masses are tuned
keeping the others fixed to study the impact on various model properties (\eg the $\CEP$). We take a 
different point of view and work out below which particular combinations of parameters determine certain features of the 
phase diagram.
\subsection{Phase border curve}
\label{sec:phasecontour}
\begin{figure}
   \centering
   \includegraphics[width=0.48\textwidth,trim = 3mm 3mm 0mm 19mm,clip=true]{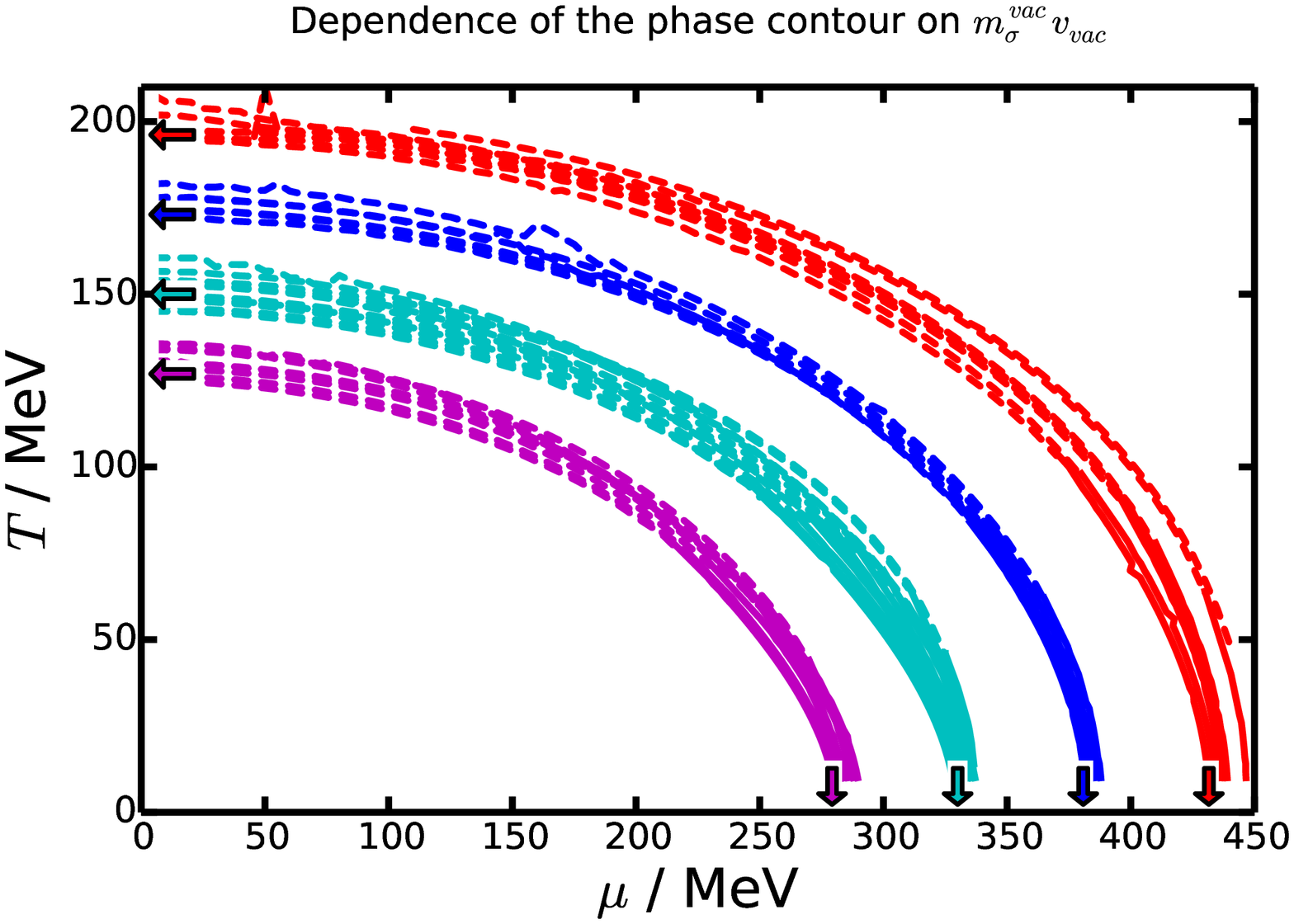}
   \caption{Critical curves (FOPT: solid, crossover estimate: dashed) for several parameter sets
   varying $\vSiVac$ in the range of $60\MeV$ to $120\MeV$ and $\mu_c^0-\mNuVac/3$ from $-100\MeV$ to $100\MeV$.
   In each group of (phase) transition curves, $\mSiVac\vSiVac$ is kept fixed at $\mSiVac\vSiVac/\GeV^2=$ 0.1156 (red),
   0.09 (dark blue), 0.0676 (light blue) and 0.048 (violet). 
   The arrows denoting the pseudocritical temperature at vanishing density, $T_\text{pc}^0$, and the critical chemical
   potential at vanishing temperature, $\mu_c^0$, are calculated according to \eqref{T_c_approx_02} for the respective 
   (same color) group of phase contours with fixed $\mSiVac\vSiVac$.}
   \label{fig:critical_curve}
\end{figure}

The phase transition features (proper FOPT curve and crossover region) are to a large extent determined by the 
meson potential at zero meson fields. This follows 
from realizing that the fermionic contribution to the pressure in the chirally restored phase is much larger
than the fermionic and bosonic contributions in the chirally broken phase. The difference is compensated at the phase
transition curve by a change in the average meson potential switching from $\langle U(\sigma\approx \vSiVac)\rangle$ 
in the chirally broken phase to $\langle U(\sigma \ll \vSiVac)\rangle$ in the chirally restored phase.
Thus 
\begin{align}
   -U(\vSiVac)+&\text{Fermi + Bose terms} \approx -U(0) +  2N_fN_c\left(\frac{7}{8}\frac{\pi^2}{90}T^4 + \frac{1}{24}\mu^2T^2 + \frac{1}{48\pi^2}\mu^4\right),\\\label{fermidruck_01}
   U(0) =& \frac{\vSiVac^2}{8}\frac{\left((\mSiVac)^2-3(\mPiVac)^2\right)^2}{(\mSiVac)^2-(\mPiVac)^2} \notag\\
        =& \frac{(\mSiVac)^2\vSiVac^2}{8}\left(1 - 5\frac{\mPiVac^2}{\mSiVac^2} + \Ord{\frac{\mPiVac^4}{\mSiVac^4}}\right),\\
   U(\vSiVac) =& \frac{(\mSiVac)^2\vSiVac^2}{8}\left(-8\frac{(\mPiVac)^2}{(\mSiVac)^2} + \Ord{\frac{\mPiVac^4}{\mSiVac^4}}\right)
\end{align}
give as an estimate for the critical temperature w.r.t. the chemical potential $T_c(\mu)$ 
\begin{align}
    T_c^2 =& \frac{1}{7\pi^2}\left(2\sqrt{30}\sqrt{\frac{42\pi^2 (U(0)-U(\vSiVac))}{2N_fN_c}+\mu^4}-15\mu^2\right).\label{T_c_approx_01}
\end{align}
Since we keep 
$(\mPiVac)^2/(\mSiVac)^2$ small in order to maintain realistic scenarios one may apply the chiral limit value of 
$U(0)-U(\vSiVac)= (\mSiVac)^2\vSiVac^2/8$ as a good estimate. Although this 
estimate looks quite crude and in the crossover region not even justified it is a surprisingly accurate 
 result for the phase transition curve (\cf Fig.~\ref{fig:critical_curve}). 
Inspecting Fig.~\ref{fig:critical_curve} one notes that although the model parameters, \eg $\vSiVac$, 
individually vary by a factor of two the form and position of the phase contour changes only slightly as long as 
$\mSiVac\vSiVac$ is kept fixed. Changing $\mNuVac$ has only small effect, too, at least if the difference between
the critical chemical potential $\mu_c^0$ at $T=0$ and the vacuum quark mass $\mNuVac/3$ is sufficiently small. (In
Fig.~\ref{fig:critical_curve} its absolute value is kept smaller than $100\MeV$.) Analyzing further parameter sets
we find that for $\mu_c^0-\mNuVac/3\gtrsim 100\MeV$ the $\CEP$ disappears, 
\cf left panel of Fig.~\ref{fig:CEP_dependence} for the dependence of the $\CEP$ temperature on this particular 
parameter combination. We are able to trace the disappearance of the $\CEP$ back to the Fermi pressure, which - for 
$\mu_c^0-\mNuVac/3$ being large enough - can compensate the difference of the meson potentials in both phases and 
thus reduces the strength of the FOPT. For $\mu_c^0-\mNuVac/3< -100\MeV$ there is the 
tendency to reduce the curvature of the phase contour, because for a higher quark mass scale, the Fermi pressure
gets less important and the pressure in the chirally broken phase is more influenced by the pressure of the pions,
which changes the $\mu$-dependence of the phase border curve.
To achieve more quantitative agreement for the pseudocritical temperature at vanishing density, $T_c^0$, and the 
critical chemical potential at zero temperature, $\mu_c^0$, it is convenient to scale the prediction according
to \eqref{T_c_approx_01} with $\mPiVac=0$ with the result for some reference parameter set. 
In Fig.~\ref{fig:critical_curve} we chose 
\begin{align}
   T_{\text{pc}}^{0,\text{ref}} = 150\MeV, && \mu_c^{0,\text{ref}} = 330\MeV && \text{ for } \mSiVac\vSiVac = 260^2\MeV^2.\label{T_c_approx_03}
\end{align}
Inspecting \eqref{T_c_approx_01} yields $T_c^0 ,\mu_c^0\propto \sqrt{\mSiVac\vSiVac}$,
thus such a scaling gives the estimates
\begin{align}
   T_{\text{pc}}^{0} \approx 150\MeV \frac{\sqrt{\mSiVac\vSiVac}}{260\MeV}, && \mu_c^0 \approx 330\MeV\frac{\sqrt{\mSiVac\vSiVac}}{260\MeV}.\label{T_c_approx_02}
\end{align}
In Fig.~\ref{fig:critical_curve} these estimates are depicted as small arrows and show good agreement with the actual
positions of the FOPT and the crossover curve.
\subsection{Isentropes}
The pattern of isentropes depends, as the $\CEP$ and the FOPT details, on the model parameters. 
Figure~\ref{fig:isentropes} (for set A)
exhibits an example where the $\CEP$ acts as an attractor for some isentropes. Such a pattern, sometimes called 
``focusing effect'' is discussed in \cite{Nonaka:2004pg,Bluhm:2006av,Nakano:2009ps} with the outcome of not being a
necessarily accompanying feature of a $\CEP$. We emphasize that isentropes provide an interesting supplementing
analytical information beyond the plain FOPT curve and the $\CEP$ position in the phase diagram.
On the FOPT curve isentropes with different $s/n$ ratios can run partially on top of each other. 
This reflects the fact that the state the model resides in is not uniquely defined on a FOPT but may differ in the 
phase decomposition. Physical properties of the medium on the FOPT curve are therefore determined as the average 
(based on \eg the volume fraction) of the respective quantity over the coexisting phases. Such a procedure is 
applied also in Figs.~\ref{fig:Compton_rate}-\ref{fig:total_rate} for the photon rates.
 
The behavior of the isentropes can be calculated analytically in the limits of $T\rightarrow 0$ as well as 
$m_q\rightarrow 0$.
In the high temperature phase, the pressure of the model is well approximated by the pressure of an ideal massless 
Fermi gas minus the meson potential at zero fields $U(\sigma=0,\pi=0)$.
For this, the entropy per baryon can be easily calculated leading to
\begin{align}
   \frac{s}{n} = \pi^2\frac{7\pi^2 \tan^3(\phi) + 15\tan(\phi)}{15\pi^2\tan^2(\phi) + 15} \label{isentr_asymp_01},
\end{align}
with $\tan(\phi) = T/\mu$. 
The meson contributions are suppressed because they acquire large masses in the hot and dense phase \cite{Scavenius:2000qd}. 
According to \eqref{isentr_asymp_01} for every choice
of $s/n$ the isentropes of an ideal massless Fermi gas, and thus for the $\LSM$ in the high temperature phase, 
follow curves with $\tan(\phi)$=const, \ie straight lines pointing to $\mu=T=0$.

\begin{figure}
   \centering
   {\includegraphics[clip=true,trim=2mm 4mm 5mm 25mm,width=0.48\textwidth]{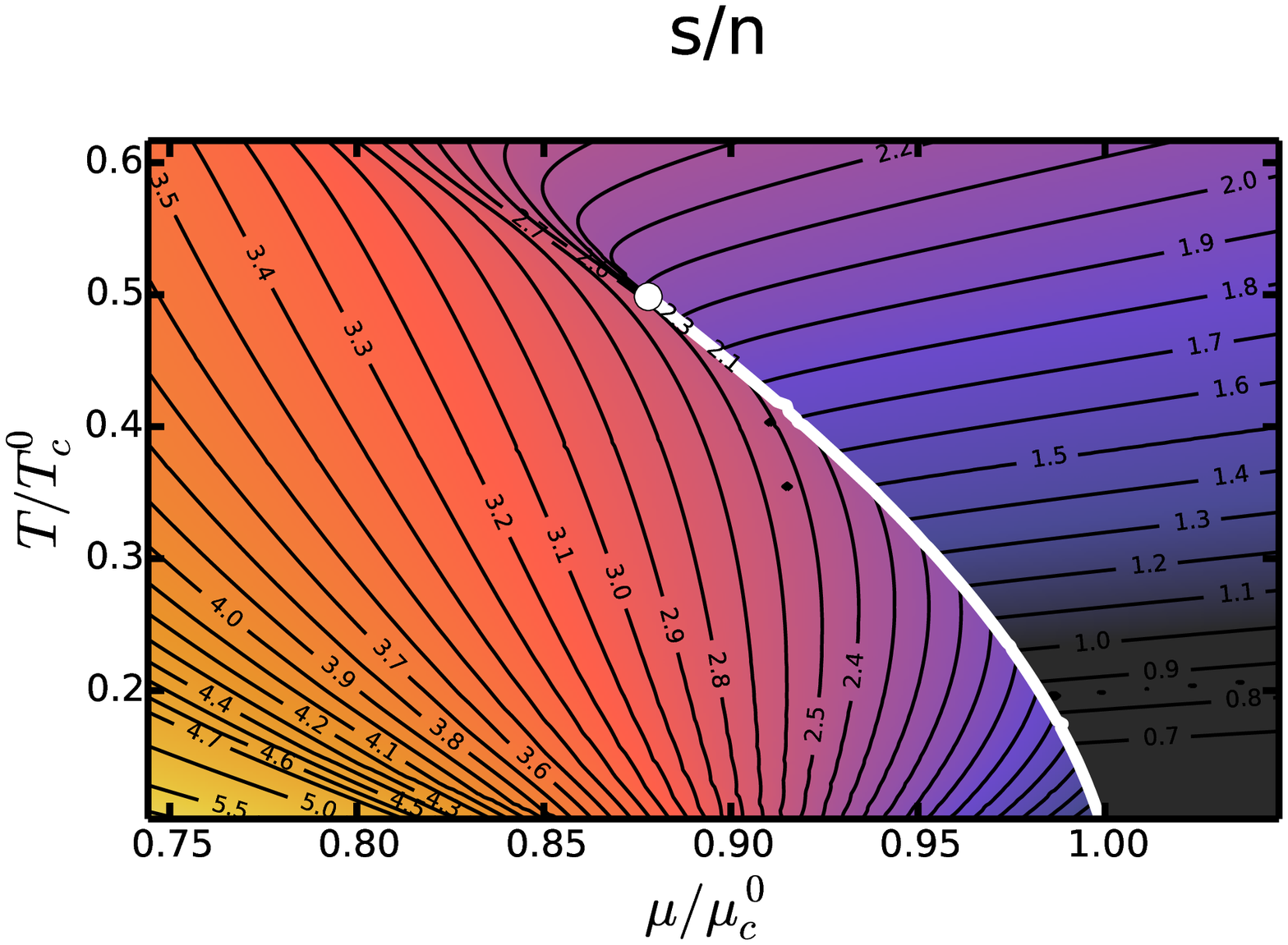}}
   {\includegraphics[clip=true,trim=2mm 4mm 5mm 25mm,width=0.48\textwidth]{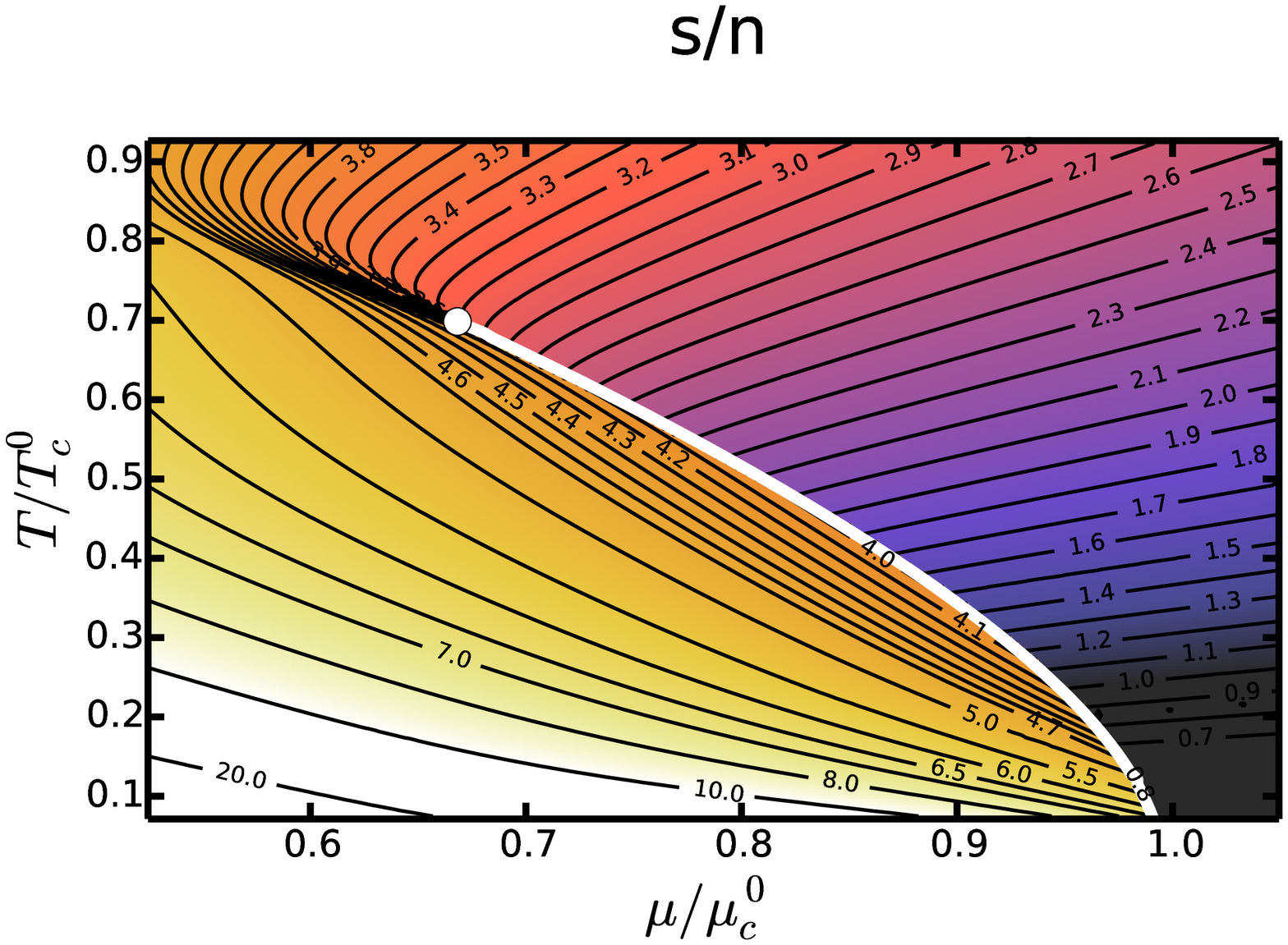}}
   \caption{Phase diagrams for parameter choices according to parameter sets 
   B (left panel) and C (right panel), \cf Tab.~\ref{tab:parameters}. $T$ is scaled by the 
   pseudocritical temperature $T_c^0$ at $\mu=0$ and $\mu$ by the critical chemical potential $\mu_c^0$ at $T=0$.
   ($T_c^0$ and $\mu_c^0$ are given in Tab.~\ref{tab:parameters}).
   For the left panel $\mNuVac/3\mu_c^0=0.907$ and for the right panel $\mNuVac/3\mu_c^0=1.111$.
   The black curves are the isentropes labeled by their corresponding ratio $s/n$.
   Shifting $m_q^\vac$ relative to $m_\sigma^\text{vac}f_\pi$ changes the dynamics of the model qualitatively.
   Definition of the FOPT curve (solid white curve) and $\CEP$ (white dot) 
            as for Fig.~\ref{fig:rate_qp_gq}.}
   \label{fig:phase_diag}
\end{figure}

The isentropes at $T\rightarrow0$ can be obtained by considering the various contributions in \eqref{ThD_pot_02} 
to the thermodynamic potential.
It turns out that the only non-vanishing term at $T=0$ in \eqref{ThD_pot_02} is the (averaged) fermion pressure 
at 
$\mu\geq m_q^\text{vac}\equiv\mNuVac/3$. Approximating the Fermi distribution function for small $T$ and $(\mu-m_q^\text{vac})$ 
one can show that all isentropes approach the point $(T=0, \mu_1=m_q^\text{vac})$ in the phase diagram, at least if 
vacuum fluctuations are not included (as in this work).
In \ref{sec:phasecontour} we discuss the dependency of the phase transition curve w.r.t. 
the model parameters finding
 that to a large extent, the critical chemical potential at $T=0$ is determined by the the combination 
$\mSiVac\vSiVac$.
Thus by tuning the model parameters (or equivalently the vacuum values for the pion and quark masses as well as $\vSiVac$, 
\cf \eqref{param_def_01}) the endpoints of the
isentropes and the FOPT curve can be shifted relative to each other making the model flexible enough for the study 
of different dynamical situations, \ie the adiabatic expansion paths are either ``going trough'', or ``sticking to''
the FOPT curve, corresponding to types IA and II in the nomenclature of \cite{Wunderlich:2016aed}. 
It turns out that within this model it is not possible by parameter tuning to shift these endpoints 
into the high-density phase.
In Fig.~\ref{fig:phase_diag} this behavior is visualized. In the left panel the isentropes approach the point
$T/T_c^0, \mu/\mu_c^0=(0, 0.907)$ which is precisely the point $(0,m_q^\vac)$ as claimed for the case that 
$m_q^\vac<\mu_c^0$. In the right panel $m_q^\vac>\mu_c^0$ and thus the isentropes all merge with the FOPT at low enough
temperatures.
\subsection{Critical end point}
To get a feeling for what determines the position of the $\CEP$ within this model one may resort to the mean field 
approximation\footnote{There is a large difference between the $\CEP$ position when comparing approximations with
and without vacuum fluctuations \cite{Nakano:2009ps}. However, when comparing approximations with and without
mesonic fluctuations the shift is much smaller and the qualitative dependence on the model parameters is similar.}. 
In this approximation the meson dependence of the pressure is only via the expectation value $v$ of the 
sigma field:
\begin{align}
   p_\text{MFA}(v) =& -\frac{\lambda}{4}(v^2-\zeta)^2 +Hv
                     -\frac{2N_fN_c T }{(2\pi)^3} 
                       \int d^3p \left[\ln\left(1+e^{(\mu-E)/T}\right) + \{\mu\rightarrow -\mu\}\right],\\
   0 =& \frac{\partial p_\text{MFA}}{\partial v} \notag\\
     =& -\lambda(v^2-\zeta)v + H + \frac{2N_fN_c g^2 v}{2\pi^2} \int dp \frac{p^2}{E} (n_F(E) + n_{\bar F}(E))\label{minimum_cond_01}
\end{align}
with $n_{F,\bar F} = (1+\exp\{(E\mp\mu)/T\})^{-1}$ denoting the distribution functions for fermions (-) and antifermions (+).
The occurrence of a FOPT and the position of the $\CEP$ are understandable
in view of \eqref{minimum_cond_01}.
A FOPT requires (at least) triple solutions of \eqref{minimum_cond_01}. One of this solutions has a small $v$ leading to a 
dominant fermion term and corresponding to the chirally restored phase. One solution is thermodynamically unstable,
and the third is relatively close to the vacuum value corresponding to the chirally broken phase. For this solution,
the derivative of the meson potential gives important contributions. 

At zero temperature, two cases can be distinguished: (i) The fermion mass $m_q = gv$ close to the critical curve (or its 
estimate according to \eqref{T_c_approx_01}) is so small that the fermionic integral in \eqref{minimum_cond_01} is 
dominant already. Then no FOPT occurs. In the opposite case, the mass can still be smaller than the 
critical chemical potential (case (iia) ) or greater (or equal) to it (case (iib) ). In both cases there is a 
FOPT. According to \eqref{T_c_approx_01} the critical curve bends toward the temperature axis
and already at relatively small $T$ the critical chemical potential is smaller than the vacuum fermion mass.
Thus we discuss only case (iib) and regard it as an upper limit for (iia). For (iib), substantial contributions
to the fermion integral origin from the edge of the Fermi distributions or their proximity, \ie the range 
$(\mu-xT,\mu+xT)$ and $x= 2\dots4$. Since the minimal argument for the Fermi distributions is $m_q$ the contributing 
interval is $[m_q, \mu+xT)$. If the vacuum quark mass is larger than $\mu+xT$ the fermion integral is too small to be of 
significance. Inserting $\mu_c(T)$ according to \eqref{T_c_approx_01} and $x=4$ evidences for $\mNuVac\gtrsim 1680\MeV$ 
(for the parameter set $\mSiVac = 700\MeV$, $\mPiVac = 138\MeV$, $\vSiVac=92.4\MeV$) that
the fermion integral is small for all temperatures at the critical curve yielding a FOPT 
surrounding the chirally broken phase completely. This provides an important observation: 
If the quark mass is sufficiently large compared to the critical chemical potential $\mu_c(T)$ whose values are 
in turn determined by the parameter combination $\mSiVac\vSiVac$,
the FOPT curve can be made to extend from the $\mu$ axis even to the $T$ axis.
On the other hand, a large fermion mass means that the isentropes end on the critical curve (see discussion above), 
which is typically not a 
desired feature and, if one needs isentropes to exit the critical curve at some non-zero  temperature, one
cannot use a parameter set with arbitrary large fermion mass, but is limited to a mass less than the critical chemical 
potential at zero temperature (determined from \eqref{T_c_approx_01}). Then, there is an upper limit for the critical 
temperature corresponding to a $\CEP$ at $T_\CEP=\Ord{100\MeV}$ and corresponding $\mu_\CEP$.

\begin{figure}
   \centering
   {\includegraphics[width=0.48\textwidth,clip=true, trim = 3mm 3mm 0mm 19mm]{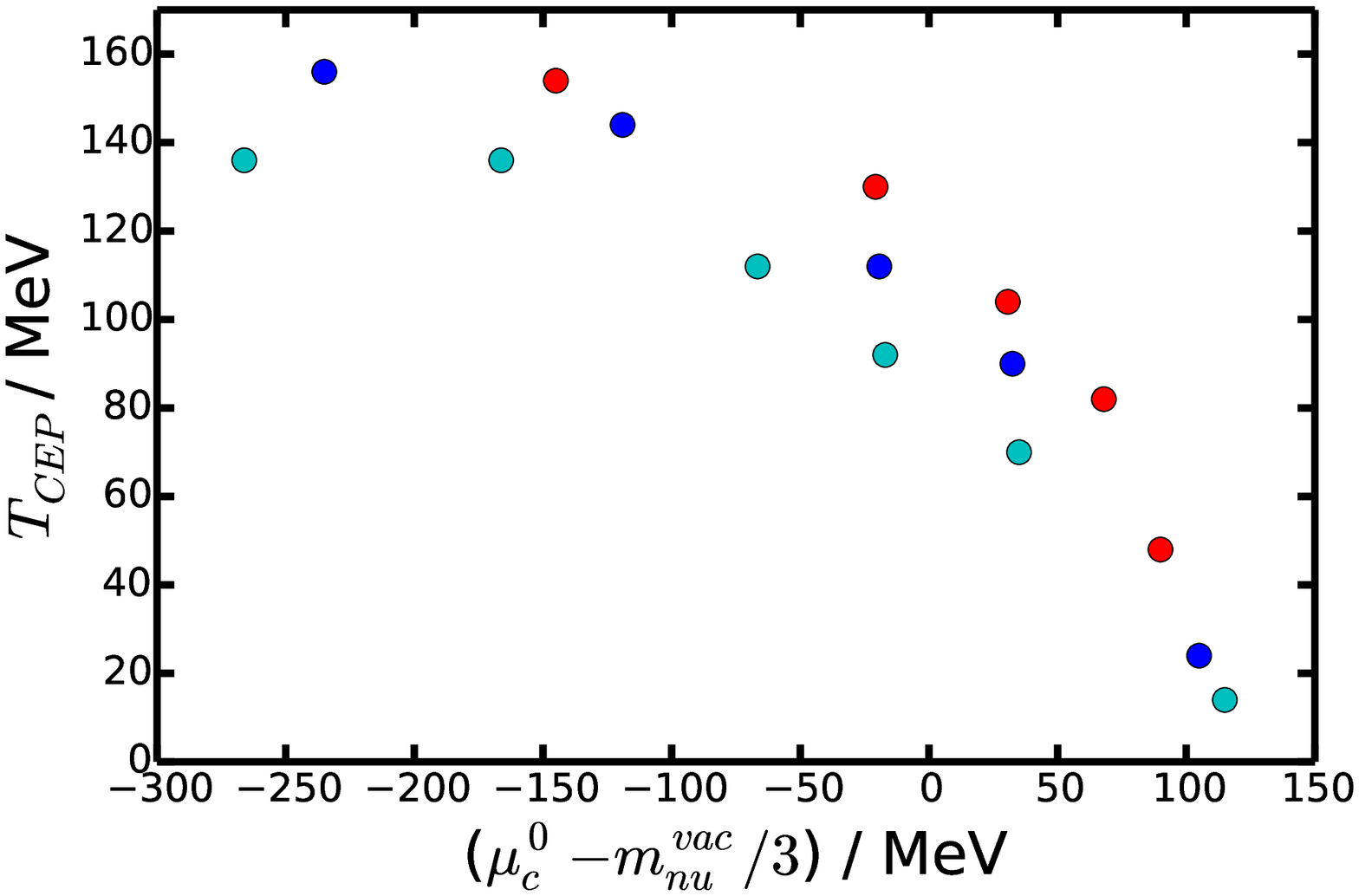}}
   {\includegraphics[width=0.48\textwidth,clip=true, trim = 3mm 3mm 0mm 19mm]{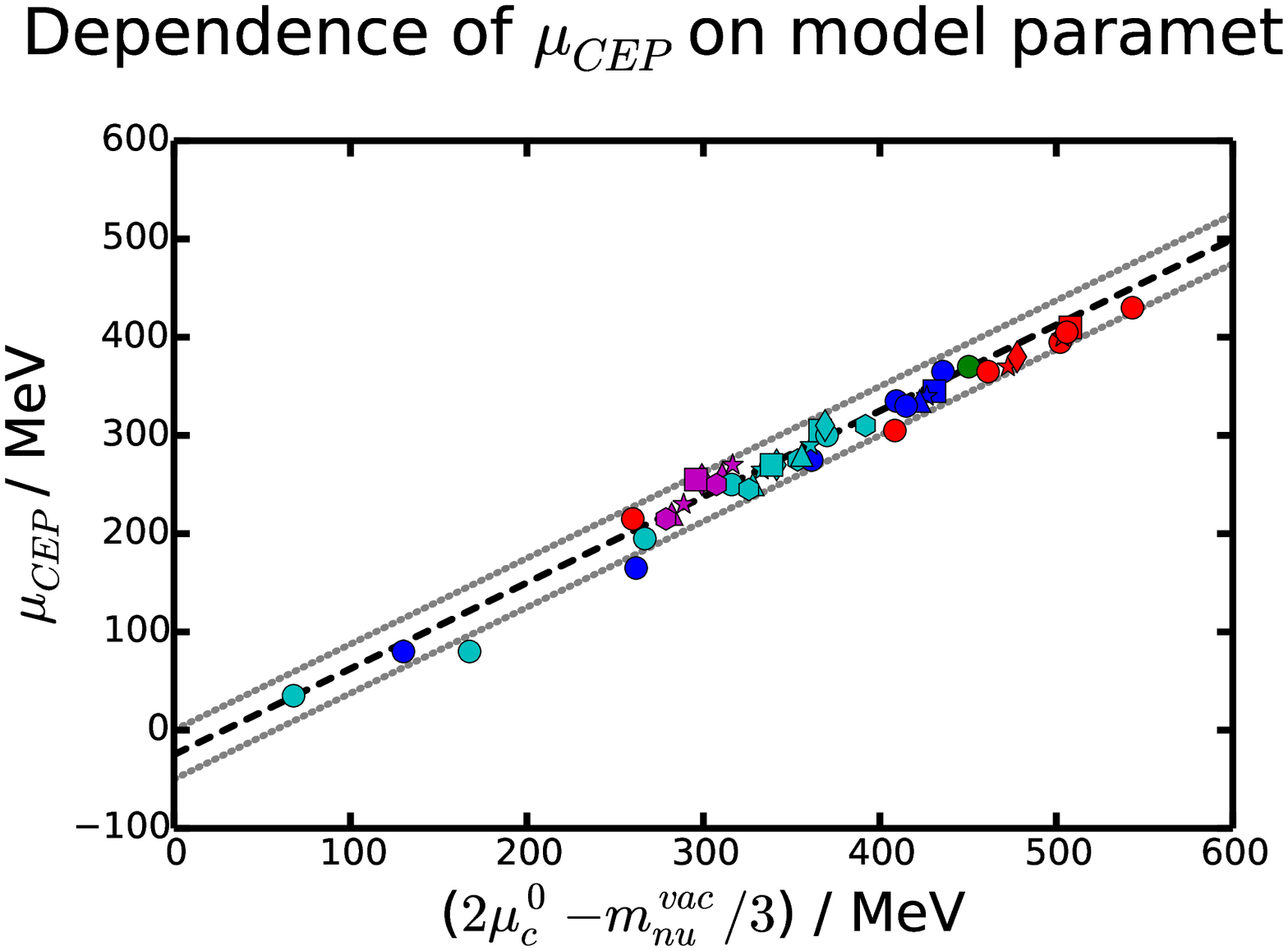}}
   \caption{Dependence of the $\CEP$ coordinates ($T_\CEP$ in the left panel and $\mu_\CEP$ in the right)
    on model parameters. The color code is the same as in Fig.~\ref{fig:critical_curve}. The symbols denote
    $\vSiVac/\MeV=$ 60 (hexagons), 70 (triangles), 90 (stars), 100 (circles), 110 (squares), 120 (diamonds). 
    The black dashed line depicts the function $f(x) = -50\MeV + 7x/8$ with $x = 2\mu_c^0-\mNuVac/3$ and the gray
    dotted lines are for $f_\pm = f(x)\pm25\MeV$.}
   \label{fig:CEP_dependence}
\end{figure}

With these considerations the behavior of the temperature $T_\CEP$ of the $\CEP$ 
(\ie increasing the vacuum fermion mass $\mNuVac/3$ increases $T_\CEP$, \cf left panel of Fig.~\ref{fig:CEP_dependence}) 
is understandable.
The chemical potentials of the various critical points collapse to one line if one assumes that the quark mass at
the phase contour is about 1/2 of its vacuum value (which works reasonable well, 
\cf right panel of Fig.~\ref{fig:CEP_dependence}).
As discussed in \cite{Ferroni:2010ct} the $\LSM$ with linearized fluctuations exhibits a fuzzy structure at the $\CEP$. 
It is therefore
more appropriate to speak of a ``$\CEP$-region'' which is hidden under the white blobs in 
Figs.~\ref{fig:Compton_rate}-\ref{fig:isentropes} and \ref{fig:phase_diag}.
Hence, we focus on the FOPT and leave the $\CEP$ related issues untouched.
\section{Matrix elements}
\label{apdx_mtrx_elem}
We quote here the matrix elements implemented in the calculations presented in Section \ref{sec:Rates_eval}.
They have been checked with the CompHEP package \cite{Pukhov:1999gg} and fulfill the corresponding Ward identities.
With the incoming momenta labeled by $p_q$ (quarks), $p_m$ (mesons) and the outgoing momenta $q_q$ (quarks) and $k$ (photon)
and the Mandelstam variables defined as $s=(p_q+p_m)^2$, $t=(p_q-q_q)^2$ and $u=(p_q-k)^2$ the fully (spin, polarization, flavor)
summed and averaged matrix elements for the Compton processes are given below.
\subsection{Compton scattering $q+\pi\rightarrow q+\gamma$}
\begin{align}
      \frac12 \sum |\mathcal{M}_{q\pi\rightarrow q\gamma}|^2 
      =& e^2 g^2\Bigg(
                  -\frac{10}{3}\left(\frac{s-m_q^2}{u-m_q^2} + \frac{u-m_q^2}{s-m_q^2}\right)
                  +\frac{20}{3}m_\pi^2 m_q^2\left(\frac{1}{(u-m_q^2)^2} + \frac{1}{(s-m_q^2)^2}\right)\notag\\
        &         +\frac43-\frac43\frac{m_\pi^2(s+u - m_\pi^2)}{(u-m_q^2)(s-m_q^2)}
                  + 8\frac{m_\pi^2t}{t-m_\pi^2}\left(\frac{1}{u-m_q^2} + \frac{1}{s-m_q^2}+\frac{2}{t-m_\pi^2}\right)
                \Bigg)\label{mtrx_elem_qpgq_01},
\end{align}
with $\sum |\mathcal{M}_{q\pi\rightarrow q\gamma}|$ denoting the spin, flavor and polarization summed matrix elements. 
The factor $1/2$ is due to averaging over incoming flavors.
For the case of massless pions (\ie in the broken phase in the chiral limit) one finds
\begin{eqnarray}
    \frac12 \sum|\mathcal{M}_{q\pi\rightarrow q\gamma}|^2 
      &=& e^2 g^2\Bigg(-\frac{10}{3}\left(\frac{s-m_q^2}{u-m_q^2} + \frac{u-m_q^2}{s-m_q^2}\right) +\frac43 \Bigg) .  
\end{eqnarray}
\subsection{Compton scattering $q+\sigma\rightarrow q+\gamma$}
\begin{eqnarray}
   \frac12 \sum|\mathcal{M}_{q\sigma\rightarrow q\gamma}|^2
      &=&  -\frac{5}{9} g^2e^2
           \Bigg((4m^2 - m_\sigma^2)\left(\frac{4m^2}{(u-m_q^2)^2}+\frac{4m^2}{(s-m_q^2)^2}+\frac{4(2m^2-m_\sigma^2)}{(u-m_q^2)(s-m_q^2)}\right)\notag\\
       &&       + 2\frac{s+7m^2-2m_\sigma^2}{u-m_q^2}+ 2\frac{u+7m^2-2m_\sigma^2}{s-m_q^2}+4 
               \Bigg)\label{mtrx_elem_qsgq_01}.
\end{eqnarray}
If the fermion masses are set to zero (corresponding to the restored phase in the chiral limit), this reduces to:
\begin{eqnarray}
   \frac12 \sum |\mathcal{M}_{q\sigma\rightarrow q\gamma}|^2
            &=&  -\frac{5}{9} g^2e^2
           \Bigg(4\frac{m_\sigma^4}{us} + 2\frac{s-2m_\sigma^2}{u}+ 2\frac{u-2m_\sigma^2}{s}+4 \Bigg).
\end{eqnarray}
\subsection{Annihilation $q+\sigma,\pi\rightarrow q+\gamma$}
The annihilation matrix elements are related to the Compton matrix elements \eqref{mtrx_elem_qpgq_01} and 
\eqref{mtrx_elem_qsgq_01} by crossing symmetries 
and can be obtained by $s\leftrightarrow t$.
The matrix elements for the anti-Compton processes \eqref{ACompton_01} are identical to those for the  
Compton processes \eqref{Compton_01}.

\bibliographystyle{aip}
\bibliography{bibliography}
\end{document}